\documentclass[preprint2]{aastex631}
\usepackage{CJK}
\received{2024 July 23}
\revised{2024 September 12}
\accepted{2024 September 22 in ApJ}
\newcommand\kmps{\mbox{km s$^{-1}$}}
\newcommand\rsun{\mbox{${\text{R}}_{\odot}$}}

\def\papertitle{Limb Observations of Global Solar Coronal Extreme-ultraviolet Wavefronts: \\
The Inclination, Kinematics, Coupling with the Expanding Coronal Mass Ejections, and Connection with the Coronal Mass Ejection Driven Shocks}

\def\paperkeywords{Solar extreme ultraviolet emission; Solar coronal mass ejection shocks; Solar coronal waves; Solar storm}
\def\papersubject{Solar: extreme ultraviolet emission; coronal mass ejection shocks; coronal waves; storm}
\hypersetup{pdfauthor=Huidong Hu (胡会东) et al.,
pdftitle=\papertitle, pdfkeywords=\paperkeywords, pdfsubject=\papersubject}

\def\affswl{State Key Laboratory of Space Weather,
    National Space Science Center, Chinese Academy of Sciences, Beijing 100190, China;
    \href{mailto:liuxying@swl.ac.cn}{liuxying@swl.ac.cn}}
\def\affnssc{Key Laboratory of Solar Activity and Space Weather,
    National Space Science Center, Chinese Academy of Sciences, Beijing 100190, China;
    \href{mailto:huhd@nssc.ac.cn}{huhd@nssc.ac.cn}}
\def\affucas{University of Chinese Academy of Sciences, Beijing 100049, China}
\def\affbjseu{Space Engineering University, Beijing 101416, China;
    \href{mailto:zhub@hgd.edu.cn}{zhub@hgd.edu.cn}}
\def\affcma{National Satellite Meteorological Center, China Meteorological Administration, Beijing 100081, China}
\def\affpku{School of Earth and Space Sciences, Peking University, Beijing 100871, China}
\def\affhusb{School of Microelectronics and Physics, Hunan University of Technology and Business, Changsha 410205, China}

\begin{document}
\begin{CJK*}{UTF8}{gbsn}
\title{\papertitle}
\shorttitle{Limb Observations of Global Solar Coronal EUV Wavefronts}
\shortauthors{Hu et al.}
\author[0000-0001-8188-9013]{Huidong Hu (胡会东)}
\affiliation{\affswl}
\affiliation{\affnssc}
\author[0000-0001-6306-3365]{Bei Zhu (朱蓓)}
\affiliation{\affbjseu}
\author[0000-0002-3483-5909]{Ying D. Liu (刘颍)}
\affiliation{\affswl}
\affiliation{\affnssc}
\affiliation{\affucas}
\author[0000-0002-2316-0870]{Chong Chen (陈冲)}
\affiliation{\affhusb}
\author[0000-0001-5205-1713]{Rui Wang (王瑞)}
\affiliation{\affswl}
\affiliation{\affnssc}
\author[0000-0002-4016-5710]{Xiaowei Zhao (赵晓威)}
\affiliation{\affcma}
\affiliation{\affpku}

\begin{abstract}
We select and investigate six global solar extreme-ultraviolet (EUV) wave events
using data from the Solar Dynamics Observatory (SDO) and the Solar and Heliospheric Observatory (SOHO).
These eruptions are all on the limb but recorded as halo coronal mass ejections (CMEs)
because the CME-driven shocks have expanded laterally to the opposite side.
With the limb observations avoiding the projection effect,
we have measured the inclination and speed of the EUV wavefront from 1.05 to 1.25 \rsun{}.
We also investigate the coupling and connection of the EUV wavefront with the CME boundary and the CME-driven shock, respectively.
The major findings in the six events are:
(1) the forward inclination of the primary and coronal-hole transmitted EUV wavefronts is estimated, respectively,
and the origins of these inclinations and their effects on the estimate of actual wavefront speed are investigated;
(2) the wavefront speed can be elevated by loop systems near the coronal base,
and the average speed in the low corona has no clear correlation with the lateral expansion of the CME-driven shock in the high corona;
(3) the fast magnetosonic Mach number of the wavefront is larger than unity from the coronal base;
(4) the EUV wavefront is coupled with the CME driver throughout the propagation in two events;
(5) after the EUV wavefront vanishes, the CME-driven shock continues traveling on the opposite side
and disconnects from the EUV wavefront in four events.
These results and their implications are discussed, which provide insight into the properties of global EUV waves.
\end{abstract}

\section{Introduction}\label{intro}
A solar extreme-ultraviolet (EUV) wave is an eruption-triggered intensity disturbance propagating in the corona,
which was first reported by \citet{ThompsonPG1998GeoRL}
in the observations of the Extreme-ultraviolet Imaging Telescope \citep[EIT,][]{DelaboudiniereAB1995SoPhEIT}
on board the Solar and Heliospheric Observatory \citep[SOHO,][]{DomingoFP1995SSRvSOHO}.
Alongside fast-mode magnetohydrodynamics (MHD) wave models
\cite[e.g.,][]{MosesCD1997SoPh,WangYM2000ApJ,GrechnevUC2011SoPh,TemmerVG2011SoPh,ZhaoWW2011ApJ},
some EUV waves are explained by non-wave models
\citep[e.g.,][]{ChenFS2005ApJ,AttrillHD2007ApJ,DelanneeTA2008SoPh}
or by a soliton \citep{WillsDaveyDS2007ApJ}.
When an EUV wave travels in all directions to a large distance comparable to the solar radius,
it is called a large-scale or global EUV wave.
A combined view incorporating both wave and non-wave components is more favored to interpret a global EUV wave.
The wave component is a leading fast-mode wave or shock,
which is driven by a laterally expanding coronal mass ejection (CME)
(the non-wave part; see \citealt{ChenWS2002ApJ,KienreichTV2009ApJ,LiuNS2010ApJ,MaRG2011ApJ,ChengZO2012ApJ,DaiDC2012ApJ,OlmedoVZ2012ApJ,ShenLS2013ApJ,LiberatoreLV2023ApJ};
and reviews by \citealt{PatsourakosV2012SoPh,Warmuth2015LRSP,Chen2016GMS,LongBC2017SoPh}).
This combined interpretation is also applicable to a small-scale EUV wave
\citep{ShenLT2017ApJ}
{as well as to some quasi-periodic fast-propagating EUV waves
\citep[e.g.,][]{SunTC2022ApJ,ZhouSZ2024SCPMA}.}
Combining the coronal plasma parameters and the kinematics,
the Mach number of an EUV wavefront and its associated shock can be estimated
\cite[e.g.,][]{WarmuthM2005AA,MaRG2011ApJ,GrechnevUC2011SoPh,CunhaSilvaSF2018AA},
which provides insight into understanding the acceleration and release of energetic particles
\citep[e.g.,][]{RouillardST2012ApJ,ParkIB2013ApJ,LarioKV2016ApJ,KouloumvakosPN2016ApJ,ZhuLK2018ApJ}.

The outermost of EUV waves in the initial stage
often has a three-dimensional (3D) dome-like shape,
which can extend to a considerable height above the solar surface
\citep[e.g.,][]{PatsourakosV2009ApJ,VeronigMK2010ApJ,LiZY2012ApJ,CunhaSilvaSF2018AA,DownsWL2021ApJ,MannV2023AA}.
The EUV wavefront heights are measured to be mostly in the range of 60--100 Mm
\citep[e.g.,][]{KienreichTV2009ApJ,PatsourakosVW2009SoPh,ShenL2012ApJ},
which may vary during the propagation
\citep[e.g.,][]{DelanneeAS2014SoPh,PodladchikovaVD2019ApJ}
or is dependent on the observation passband
\citep[e.g.,][]{HouTW2022ApJ}.
Stereoscopic analysis based on multiple-viewpoint imaging
has been applied to determine the height
\citep[e.g.,][]{KienreichTV2009ApJ,PatsourakosVW2009SoPh,DelanneeAS2014SoPh,PodladchikovaVD2019ApJ}.
Due to the effects of projection or line-of-sight integration,
estimate of the height of the EUV wavefront is still challenging,
which can also affect the measurement of the wavefront kinematics
\citep{KienreichTV2009ApJ,DownsWL2021ApJ}.
Models and observations suggest that the 3D EUV wavefront in the corona
is forward inclined in the propagation direction
\citep[e.g.,][]{UchidaY1968SoPh,AfanasyevU2011SoPh,LiuON2012ApJ,LiuJD2018ApJ,HouTW2022ApJ}.
The actual speed direction of an EUV wavefront is perpendicular to the wavefront,
and is toward the solar surface in the low corona.
Therefore, the actual wavefront speed is smaller than the observed apparent speed.
The wavefront inclination angle can hardly be determined without limb observations even from multiple perspectives.
Investigations of EUV waves started on the solar limb
are necessary to obtain the actual kinematics of EUV waves.

An EUV wave may decouple from the driving CME
when the lateral expansion of the CME decelerates or ceases,
after which the leading wave component travels freely
\citep[e.g.,][]{CohenAM2009ApJ,PatsourakosV2009ApJ,ChengZO2012ApJ,DaiDC2012ApJ,LiberatoreLV2023ApJ}.
The wave component is often a diffuse bright front followed by the bright CME boundary
\citep[e.g.,][]{CohenAM2009ApJ, ChengZO2012ApJ, DaiDC2012ApJ}
or by the expanding dimming CME bubble
\citep{PatsourakosV2009ApJ}.
The decoupling usually occurs during the propagation,
which means that before fading out the EUV wavefront
travels a distance greater than the lateral width of the CME.
For a large-scale CME that expands to a broad angular width,
it is still unknown whether the CME and EUV wavefront are coupled throughout the wavefront propagation.

A CME-driven shock observed in white light in the outer corona
is often connected with an EUV wave, the footprint of the shock in the low corona
\citep[e.g.,][]{VeronigMK2010ApJ,ChengZO2012ApJ,KwonZO2014ApJ,ZhuLK2018ApJ,LiuZZ2019ApJ}.
This connection indicates the shock nature of the corresponding EUV wave and the global geometry of the shock.
{Most solar type II radio bursts are signatures of CME-driven shocks
\citep[e.g.,][]{UchidaY1960PASJ,MannCA1995AA,ClassenA2002AA,LiuLB2009ApJ,ChoBM2011AA,KumariMK2023AA},
and some are attributed to non-CME generated coronal shocks
\citep[e.g.,][]{MagdalenicMZ2010ApJ,SuCD2015ApJ,HouTS2023ApJ}.
For a CME-driven shock, a type II radio burst can be used to infer the shock's propagation properties}
\citep[e.g.,][]{ReinerKB2007ApJ,LiuLM2008ApJ,LiuLL2013ApJ,CremadesIS2015SoPh,HuLW2016ApJ,ZhaoLH2019ApJ}.
A CME-driven shock can propagate laterally to the other side of the Sun opposite the eruption site,
which has been observed with white light in the outer corona
\citep{KwonZO2014ApJ,LiuHZ2017ApJ,HuLZ2019ApJ}.
Henceforth, for the sake of brevity,
we refer to a ``CME-driven shock observed in white light'' as a ``white-light shock''.
In the contribution of the transmission of two polar coronal holes,
the EUV wave on 2017 September 10 travels all over the solar sphere
and reaches the opposite side of the eruption
\citep[e.g.,][]{LiuJD2018ApJ,HuLZ2019ApJ}, which is unique.
Although the eruption is on the west limb,
the CME manifests as a halo because the bubble of the white-light shock encloses the Sun
\citep[similar to cases in][]{KwonZO2014ApJ}.
The wavefront of this EUV wave, throughout the propagation,
is continuously spatially connected with the white-light shock even on the opposite side.
However, in common events where the CME-driven shock propagates to the opposite side
without the transmission of large-scale coronal holes,
the connection between the EUV wave and the white-light shock on the opposite side is still elusive.

In this work, we will select and investigate six global EUV wave events which are all triggered by eruptions on the solar limb.
The associated CMEs are all recorded as halos
because their white-light shocks have propagated to the other side of the Sun opposite the eruption site.
This work aims to comprehensively study (1) the local morphology and (2) kinematics of global EUV waves in the low corona,
as well as (3) the coupling with the expanding CMEs and (4) connection with the CME-driven shocks.
Observations of limb events can avoid the projection effect,
which makes it simpler to obtain the inclination and kinematics of the EUV wavefront.
Limb observations are also effective to determine the CME boundary, the EUV wavefront, and the CME-driven white-light shock,
which is helpful to discern the decoupling of the EUV wavefront from the CME,
as well as the connection between the EUV wavefront and the white-light shock.
The data, event selection, and the overview of the events are presented in Section \ref{data}.
The analysis results are provided in Section \ref{results} and are discussed in Section \ref{discussions}.
The conclusions are remarked in Section \ref{conclusions}.

\begin{deluxetable*}{cccclccc}
\tablenum{1}
\tablecaption{List of six EUV wave events associated with halo CMEs from the Solar Limb \label{tablist}}
\tablewidth{0pt}
\tablehead{
\colhead{Event Number} & \colhead{Location} & \colhead{Flare} & \colhead{Side} &
\colhead{Inclination} & \colhead{Speed} & \colhead{Shock Exp.} & \colhead{CH?}  \\
\colhead{(YYYY-MM-DD)} & \colhead{(\degr)} & \colhead{} & \colhead{} & \colhead{(\degr)} &
\multicolumn{2}{c}{(\kmps)} & \colhead{}
}
\colnumbers
\startdata
2011-09-22 & N09E89  & X1.4   & Southward  & 64$\pm$3             & 333$\pm$18  & 1526$\pm$75  & No  \\
2013-05-13 & N11E90  & X1.7   & Northward  &    ---               & 451$\pm$15  &  811$\pm$49  & No  \\
2013-11-07 & S11W97  & M1.8   & Southward  & 65$\pm$2             & 336$\pm$6   &  680$\pm$33  & No  \\
2015-05-05 & N15E80  & X2.7   & Northward  & 66$\pm$1 (38$\pm$5)  & 362$\pm$21  &  823$\pm$40  & Yes \\
2017-09-10 & S09W92  & X8.2   & Southward  & 75$\pm$4 (38$\pm$1)  & 772$\pm$85  & 1930$\pm$202 & Yes \\
2020-11-29 & S23E97  & M4.4   & Northward  & 61$\pm$1             & 468$\pm$15  & 1527$\pm$122 & No  \\
\enddata
\tablecomments{Column (4) indicates which side of the EUV wavefront is investigated.
Column (5) lists the average inclination angles of the primary wavefront,
whose average range is marked by the two vertical dotted lines
in panel (a) of Figures \ref{k20110912}, \ref{k20131107} -- \ref{k20201129};
in the brackets are the average inclination angles of the transmitted wavefronts (see the text),
of which the average range is given by the two vertical dashed lines
in panel (a) of Figures \ref{k20150505} and \ref{k20170910};
the uncertainty is the standard error of the mean.
Column (6) gives the average speed of the primary wavefront,
which is from averaging over the time span denoted by the horizontal solid line
in panel (a) of Figures \ref{k20110912} -- \ref{k20201129}
and then over altitudes of 1.05 -- 1.25 {\rsun}.
Column (7) provides the lateral-expansion speed of the CME-driven shock fitted with the ellipsoid model (see the text).
Column (8) indicates whether the EUV wave is transmitted by a coronal hole (CH) on the limb during the propagation.}
\end{deluxetable*}

\section{Data and Events}\label{data}
All six selected EUV events are associated with halo CMEs recorded in the SOHO/Large Angle and Spectrometric Coronagraph Experiment (LASCO) CME catalog
\citep{GopalswamyYM2009EMP},
which are all initiated by eruptions on the solar limb from the view of the Earth.
Because high-cadence data are better to acquire the kinematics of EUV waves, especially for those fast ones
\citep{LongGM2008ApJ,PatsourakosV2012SoPh,NittaST2013ApJ,DownsWL2021ApJ},
we use data of the Atmospheric Imaging Assembly \citep[AIA,][]{LemenTA2012SoPh}
on board the Solar Dynamics Observatory \citep[SDO,][]{PesnellTC2012SoPh} with a cadence of 12 seconds.
Although the EUV wave is prominent in running-difference images of AIA 211, 193, and 171 {\AA},
the 211 {\AA} images are chosen because they have a relatively better feature of the EUV wavefront in our events.
Below are the selection criteria for the events to be studied:
(1) Only halo CMEs with a total width of 360\degr{} noted in the CME catalog are selected,
which means that each CME-driven shock has propagated to the other side opposite the eruption site;
(2) Events with eruption sites at heliographic longitudes between 80\degr{} and 100\degr{} are considered to be limb events in this study;
(3) On the limb an EUV wave has southward and northward wavefronts,
of which the wavefront with a larger propagation distance is chosen for investigation in each event;
(4) If both the southward and northward wavefronts are too diffuse to be identified
or have a propagation distance less than 60\degr{} from the eruption site in the position angle,
the corresponding events are also excluded.
With these criteria, six events have been selected from dozens of limb events from 2011 September to 2020 November,
and are listed in Table \ref{tablist}.
Most of these limb events are remarkable and have been studied in prolific literatures.
Below in this section we will give an overview for each event, and the detailed results will be presented in the following sections.

The 2011 September 22 EUV wave event:
This event is associated with an X1.4 flare peaked at 11:01 UT on 2011 September 22
at N09E89 in active region (AR) 11302 on the east limb.
The southward EUV wavefront is included as a sample in our study,
which is marked in Column (4) of Table \ref{tablist}.
The flare and the evolution of the erupting flux rope have been studied with observations of multiple wavelengths
\citep[e.g.,][]{AkimovBM2014MNRAS,ZhangTL2023ApJ}.
A shock is formed by the lateral expansion of the CME, which is signified by type II radio bursts
\citep[e.g.,][]{CarleyLB2013NatPh,ZuccaCB2014AA} and by intensive flux of solar energetic particles (SEP)
\citep[e.g.,][]{GopalswamyXA2014EPS}.

The 2013 May 13 EUV wave event:
This event is accompanied by an X1.7 flare peaked at 02:17 UT on 2013 May 13
at N11E90 in AR 11748 on the east limb.
The more prominent northward EUV wavefront is selected for our study.
High intensity of SEP is observed in this event
\citep[e.g.,][]{GopalswamyXA2014EPS,ParkIB2015ApJ,AschwandenCC2017ApJ},
and type II radio bursts associated with this eruption are also detected
\citep[e.g.,][]{RichardsonRC2014SoPh,ShareMW2018ApJ}.
These suggest that a CME-driven shock is produced by this eruption.

The 2013 November 7 EUV wave event:
This event is associated with an M1.8 flare peaked at 00:02 UT on 2013 November 7 in AR 11890.
The southward wavefront is taken into analysis for our study.
The eruption site is at S11W97 slightly behind the west limb,
which indicates that the actual flare intensity is probably higher than the observed.
A short duration of type II radio burst was detected around 00:30 UT by the Wind spacecraft
\citep{BougeretKK1995SSRv} and by both the twin STEREO spacecraft
\citep{BougeretGK2008SSRv} (the data are not shown here),
which provides evidence of a CME-driven shock in this event.

The 2015 May 5 EUV wave event:
This event is along with an X2.7 flare peaked at 22:11 UT on 2015 May 5 in AR 12339
\citep{MilliganHC2020SpWea,ZimovetsNS2022GeAe},
which also has a white-light flare
\citep{NamekataSW2017ApJ}.
The eruption is at N15E80 on the east limb,
which is related to a CME-driven shock associated with type II radio bursts
\citep{JohriM2016SoPh}.
The northward wavefront is more prominent and is selected for our study.

The 2017 September 10 EUV wave event:
This is a star textbook event, which has been studied in numerous publications.
The eruption is at S09W92 in AR 12673 on the west limb,
associated with an X8.2 flare peaked at 16:06 UT on 2017 September 10.
The flare has been investigated by ground- and space-based instruments at multiple wavelengths
\citep[e.g.,][]{SeatonD2018ApJ,OmodeiPL2018ApJ,GaryCD2018ApJ,HayesGD2019ApJ,ZhaoLV2021ApJ}.
Turbulent magnetic reconnections are observed in an extended current sheet,
which have been studied in a plethora of observations or models
\citep[e.g.,][]{YanYX2018ApJ,LiXD2018ApJ,ChengLW2018ApJ,FrenchJM2019ApJ,ChenSG2020NatAs,YuCR2020ApJ}.
The CME and its driven shock are both of spectacular speed and scale
\citep[e.g.,][]{GopalswamyYM2018ApJ,VeronigPD2018ApJ,HuLZ2019ApJ,LiuZZ2019ApJ,YangWF2021ApJ}.
The CME-driven shock is also related to type II radio bursts and contributes to the acceleration of energetic particles
\citep[e.g.,][]{ZhaoLC2018RAA,GopalswamyYM2018ApJ,MorosanCH2019NatAs,ZhuLK2021ApJ,deKoningPS2022ApJ}.
A series of interplanetary sequences of this event have been observed
\citep[e.g.,][]{LeeJL2018GeoRL,CramerWE2020JGRA,DingLH2020RAA,KocharovPL2020ApJ}.
The EUV wave is currently the only known one that travels all over the solar sphere
in the contribution of the transmission of two polar coronal holes
\citep{LiuJD2018ApJ,HuLZ2019ApJ}.
In this study, we will investigate the local geometry and kinematics of the EUV wavefront in the low corona,
as well as the connection with the associated CME and white-light shock.
The southward wavefront is analyzed because it is more prominent than the northward wavefront in AIA 211 {\AA} images.

The 2020 November 29 EUV wave event:
This event is accompanied by an M4.4 flare peaked at 13:11 UT on 2020 November 29 in AR 12790,
which triggers the first widespread SEP event of solar cycle 25 detected by multiple recently launched spacecraft
\citep[e.g.,][]{CohenCC2021AA,KollhoffKL2021AA,FuDZ2022ApJ}.
A white-light CME-driven shock and its associated type II radio burst are also observed
\citep[e.g.,][]{KollhoffKL2021AA,ChenLZ2022ApJ,NievesChinchillaAC2022ApJ}.
The flare is at S23E97 on the east limb, which is partially occulted, similar to the 2013 November 7 event,
and the actual flare intensity is also possibly higher than the observed.
As noted in Table \ref{tablist}, the northward wavefront is selected for our study.

All the six EUV wave events are initiated on the solar limb, but are all observed as full halo CMEs,
where the 360\degr{} ``halo'' is contributed by the white-light CME-driven shock that propagates to the opposite side in the corona.
We will reveal the inclination and kinematics of the EUV wavefronts in the low corona,
and discuss the coupling with the expanding CMEs and the connection with the white-light shocks in the following sections.

\begin{figure*}[ht!]
\centering
\includegraphics[width=0.81\textwidth]{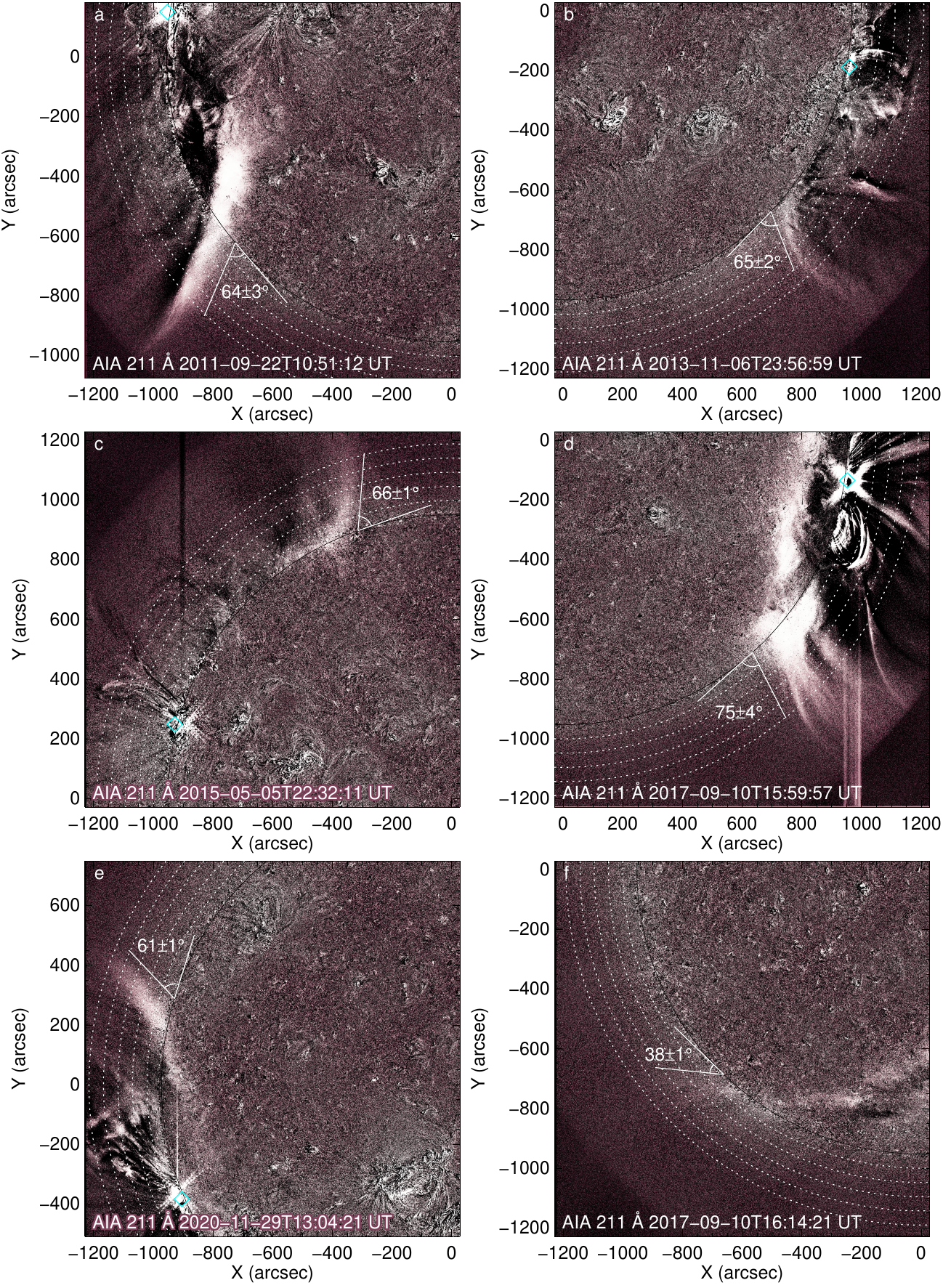} 
\caption{\label{stillMaps}{Still running-difference EUV images of SDO/AIA 211 {\AA}
for five of the selected six events except the 2013 May 13 event.
(a)--(e) show clear inclination of the primary wavefront.
(f) displays the forward inclination of the transmitted wavefront in the 2017 September 10 event.
The angles in units of degrees are the average inclination angles of the wavefronts over specified time periods
(see the note of Table \ref{tablist} and the text).
The cyan diamonds mark the eruption sites.
The dashed-circles denote the heights from 1.05 to 1.25 {\rsun} with an increment of 0.05 {\rsun}.}}
\end{figure*}

\begin{figure*}[ht!]
\centering
\includegraphics[width=\textwidth]{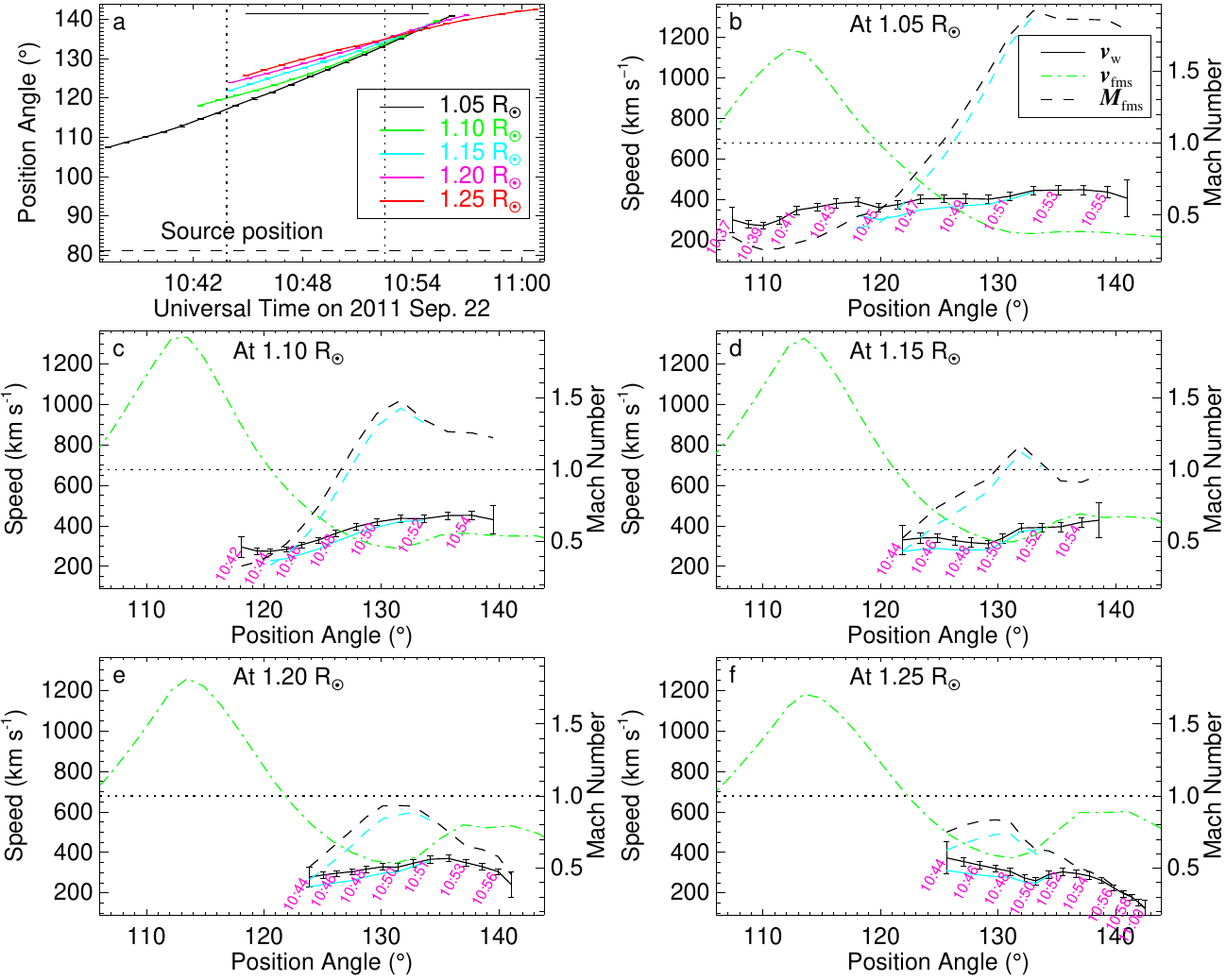}
\caption{\label{k20110912}{Time-distance profiles in position angles
(measured in degrees counterclockwise from to the solar North) ((a))
and kinematics ((b)--(f)) of the EUV wavefront,
with an interval of 1 minute at altitudes from 1.05 to 1.25 {\rsun} for the 2011 September 22 event.
In panel (a),
the two vertical dotted lines indicate the time period during which the inclination angle is measured and averaged;
the horizontal solid line marks the time span
used to calculate the average speed in Column (6) of Table \ref{tablist} and Figure \ref{vles};
the horizontal dashed line denotes the position angle of the source region of the EUV wave (i.e., the eruption site).
In panels (b)--(f),
the solid curve represents the apparent wavefront speed ($v_{\text{w}}$)
that is the multiplication of the derivative of the time-distance profile in (a) (converted to radians) and the corresponding altitude,
where the derivative is computed by the \texttt{DERIV.pro} function in the Interactive Data Language (IDL);
the cyan solid curve is the actual speed calculated by multiplying the apparent speed with the sine of the inclination angle;
the green dash-dotted curve indicates the fast magnetosonic speed ($v_{\text{fms}}$)
given by an MHD model at each altitude (see the text);
the black and cyan curves are the fast magnetosonic Mach number ($M_{\text{fms}} = v_{\text{w}} / v_{\text{fms}}$) scaled by the right axis,
corresponding to the apparent and actual wavefront speeds, respectively;
the pink text labels the approximate Universal Time for each closest data point.}}
\end{figure*}

\begin{figure*}[ht!]
\centering
\includegraphics[width=\textwidth]{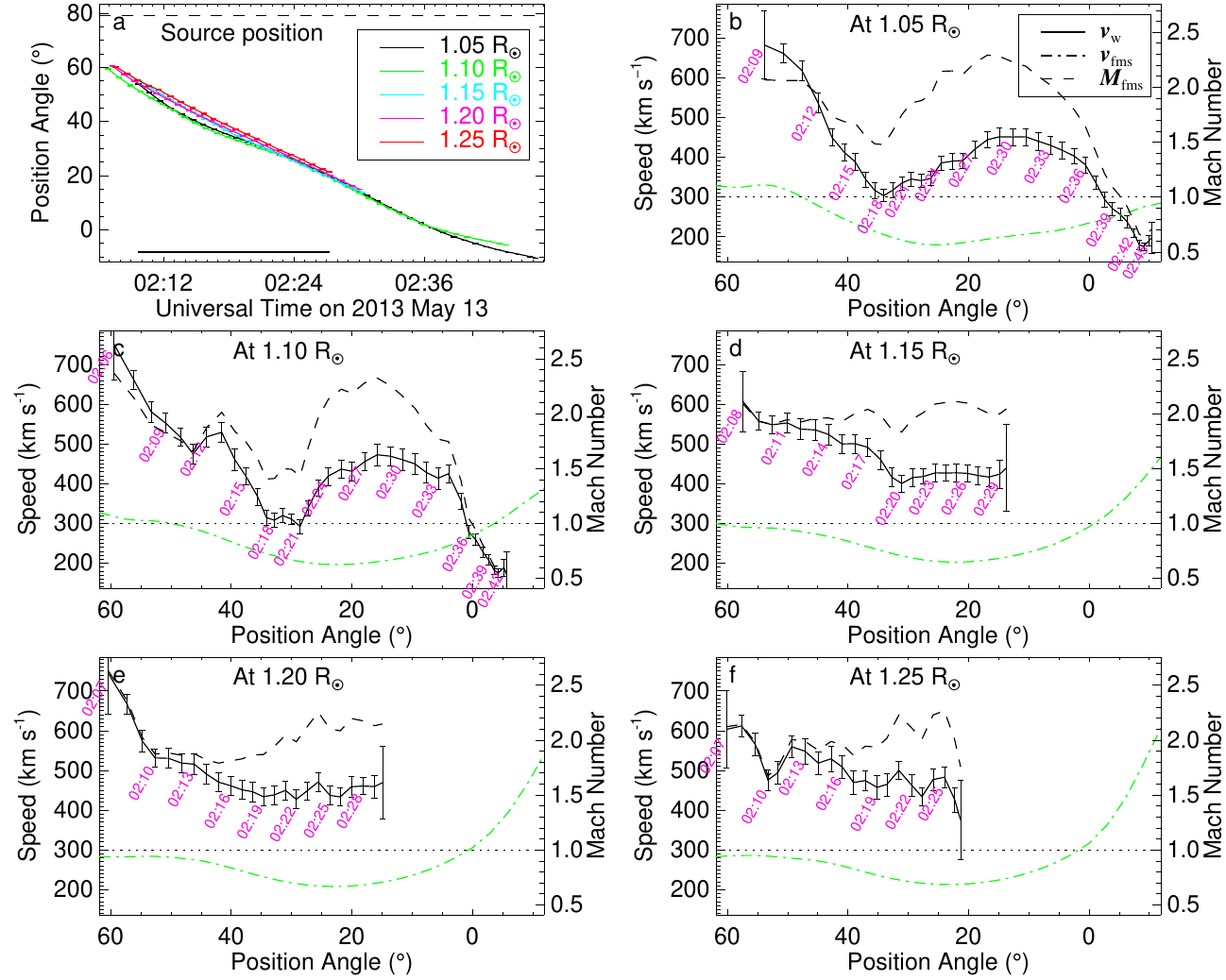}
\caption{\label{k20130513}{Similar to Figure \ref{k20110912}, but for the 2013 May 13 event.
The difference from Figure \ref{k20110912} is that there are no types of lines related to the wavefront inclination.}}
\end{figure*}

\begin{figure*}[ht!]
\centering
\includegraphics[width=\textwidth]{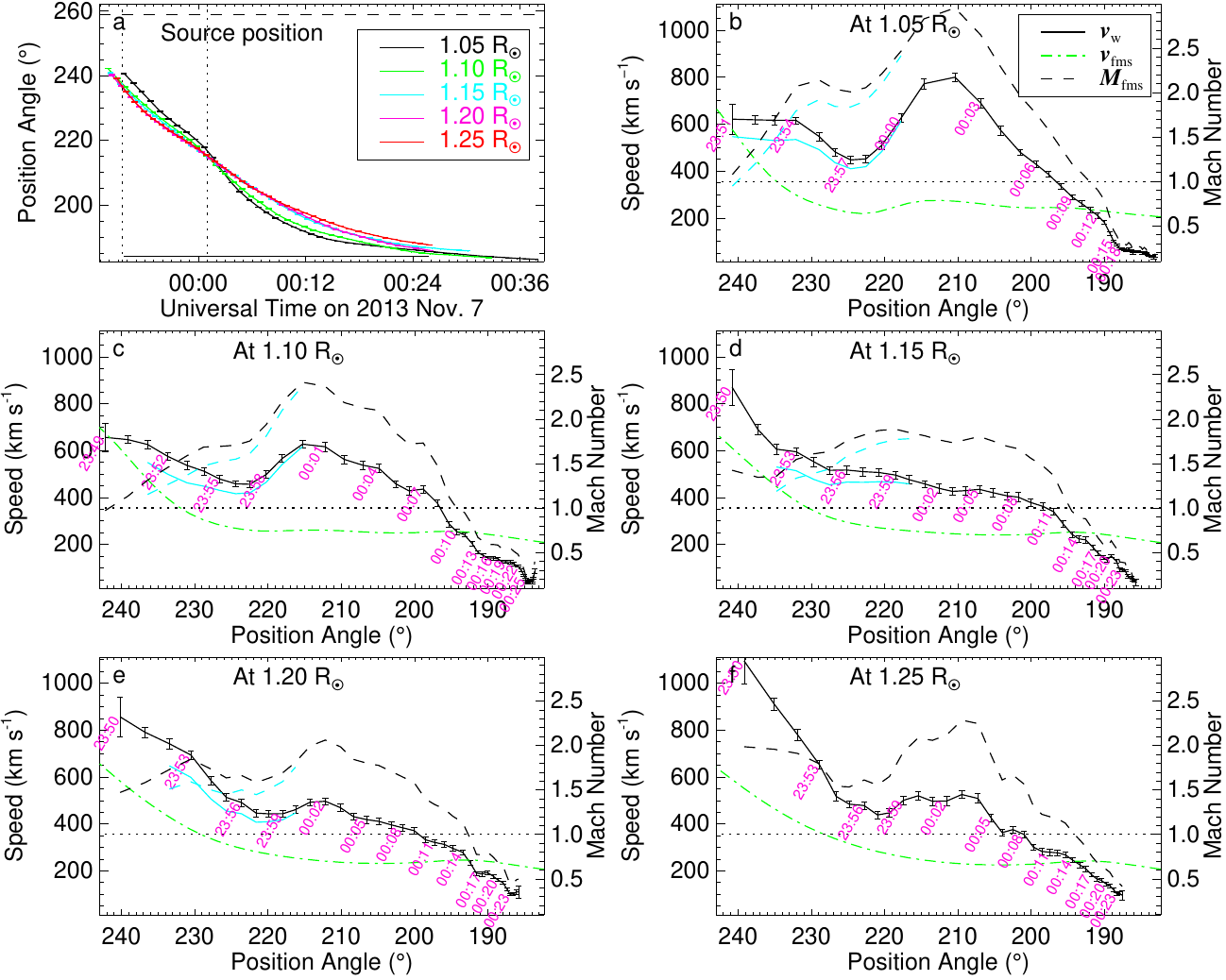}
\caption{\label{k20131107}{Similar to Figure \ref{k20110912}, but for the 2013 November 7 event.}}
\end{figure*}

\begin{figure*}[ht!]
\centering
\includegraphics[width=\textwidth]{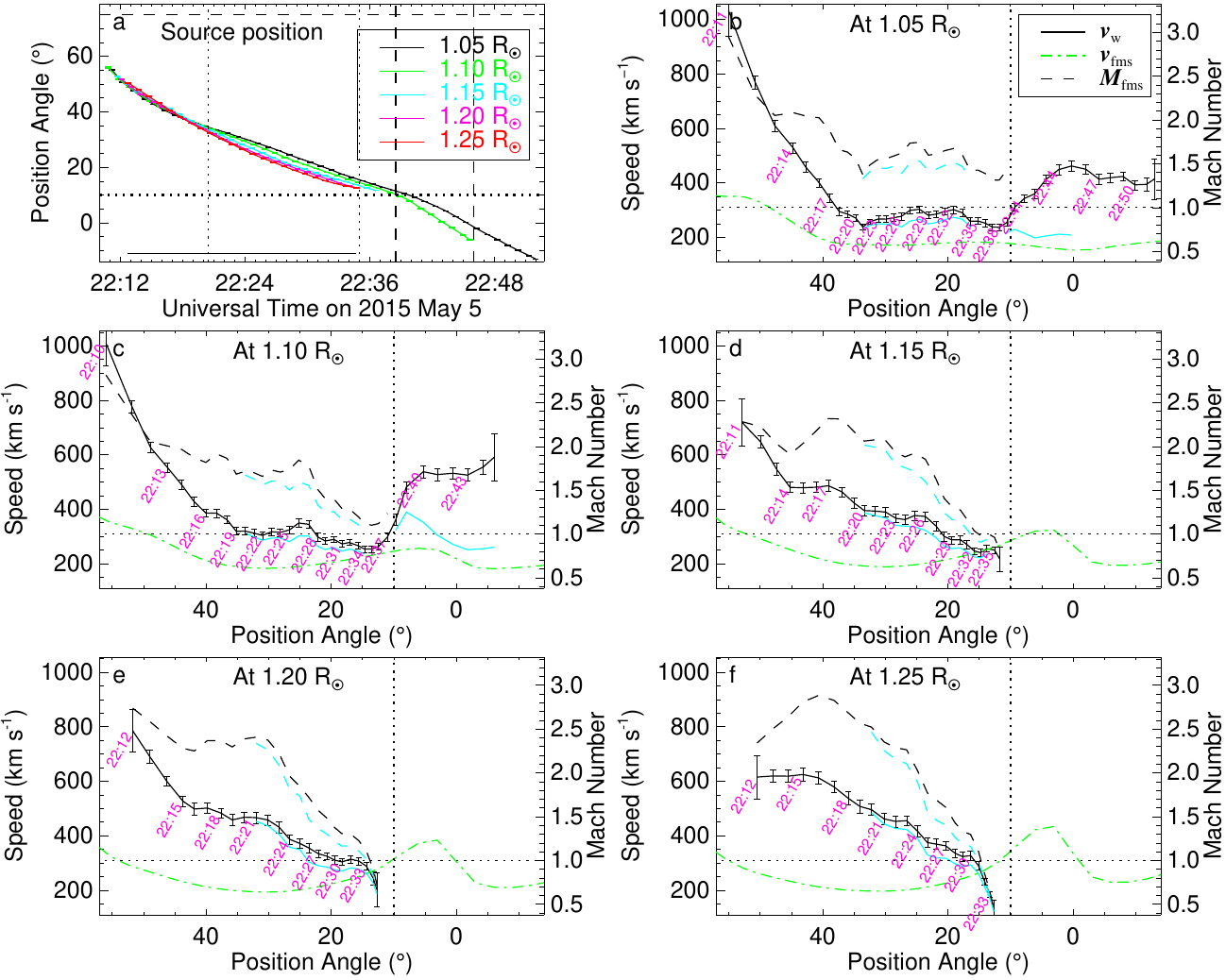}
\caption{\label{k20150505}{Similar to Figure \ref{k20110912}, but for the 2015 May 5 event.
The additional thick dotted line (horizontal in panel (a) and vertical in panels (b)--(f))
marks the approximate position of the boundary of a north polar coronal hole, from where the EUV wave is transmitted.
The two vertical dashed lines in panel (a) indicate the time period
in which the inclination angle of the transmitted wavefront is estimated and averaged
(see Table \ref{tablist} and the text).}}
\end{figure*}

\begin{figure*}[ht!]
\centering
\includegraphics[width=\textwidth]{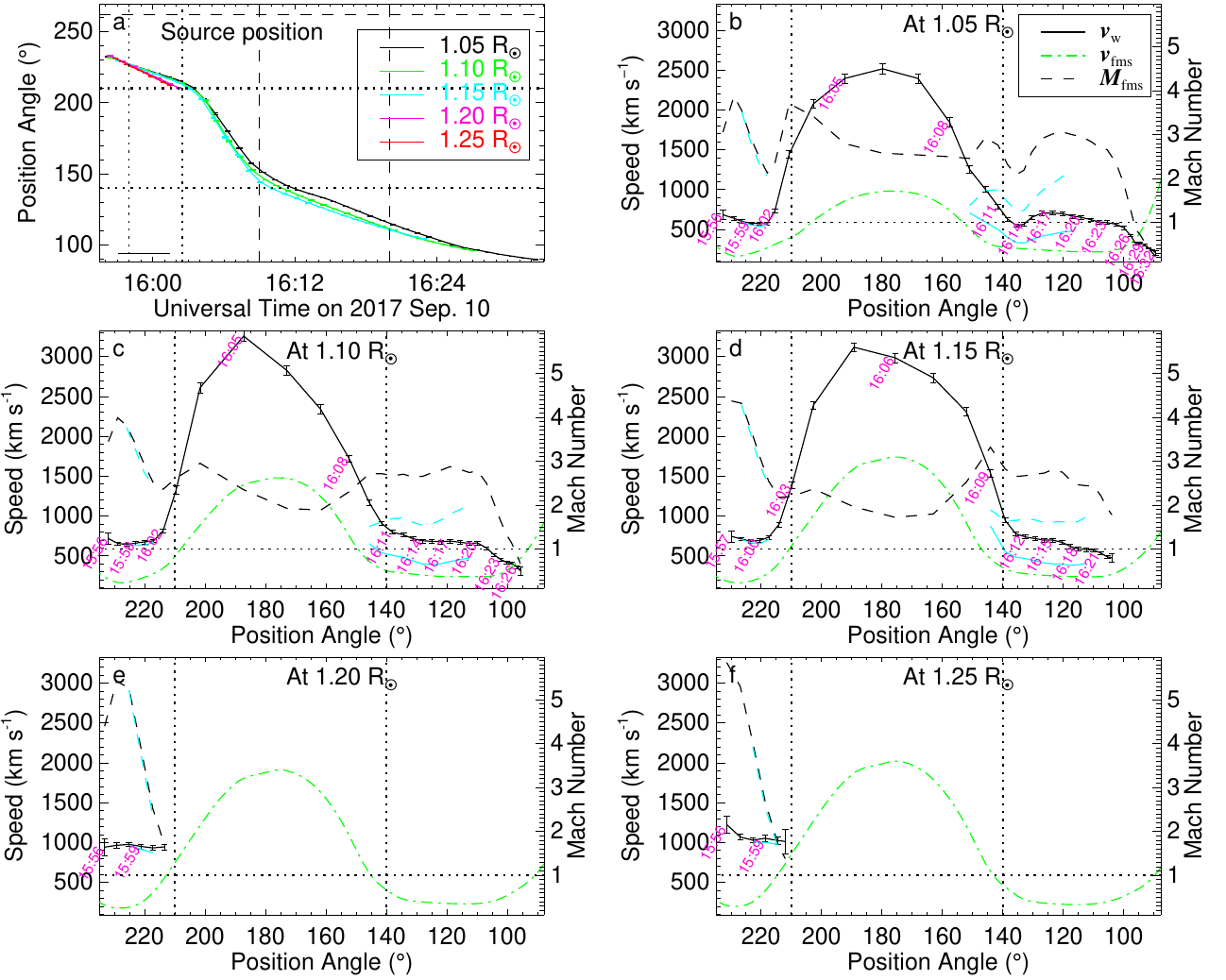}
\caption{\label{k20170910}{Similar to Figure \ref{k20110912}, but for the 2017 September 10 event.
The additional thick dotted lines (horizontal in panel (a) and vertical in panels (b)--(f))
indicate the boundaries of a south polar coronal hole where the EUV wave is transmitted.
The two vertical dashed lines in panel (a) mark the time period
where the inclination angle is measured and averaged for the wavefront after being transmitted by the coronal hole.}}
\end{figure*}

\begin{figure*}[ht!]
\centering
\includegraphics[width=\textwidth]{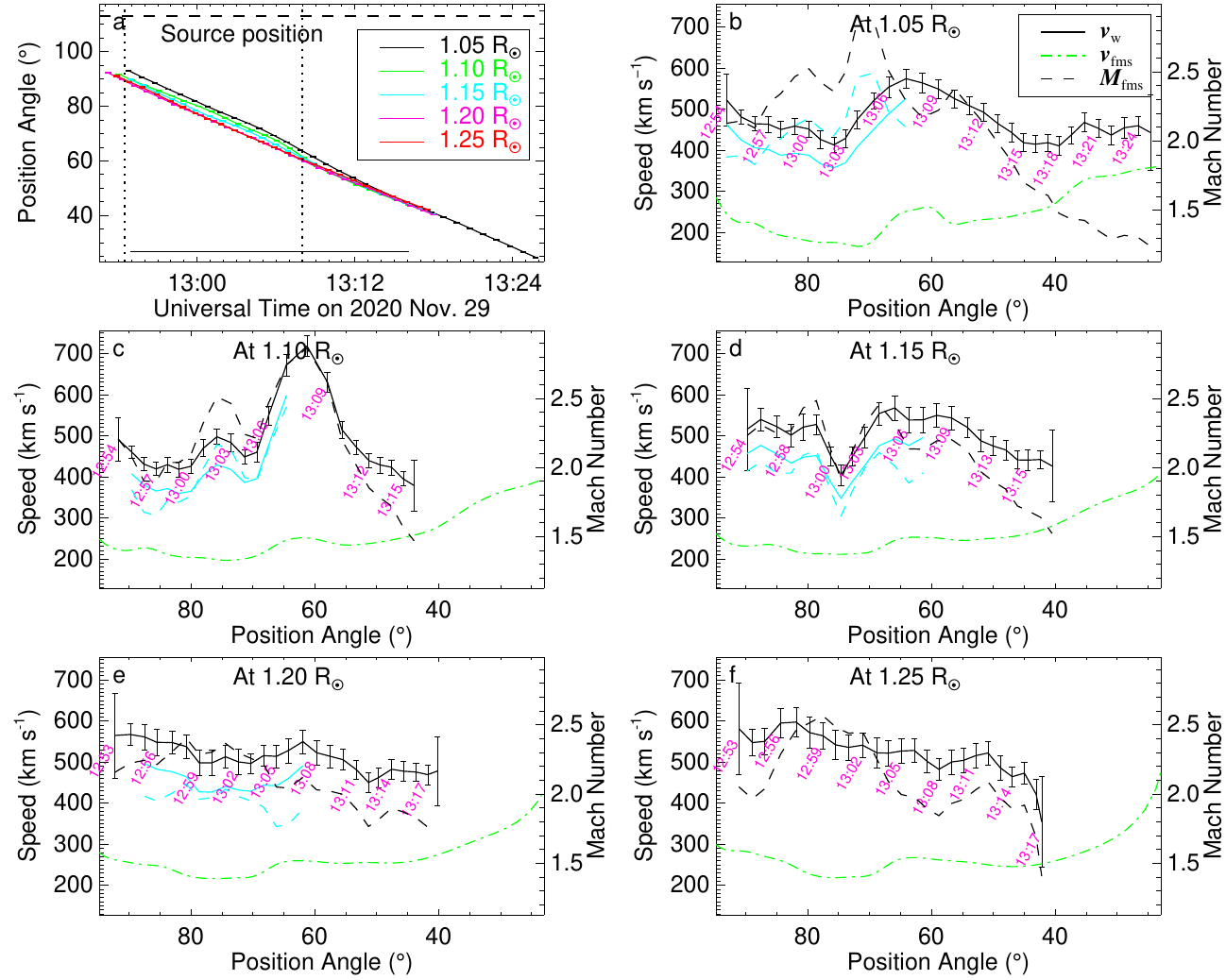}
\caption{\label{k20201129}{Similar to Figure \ref{k20110912}, but for the 2020 November 29 event.}}
\end{figure*}

\section{Analysis and Results}\label{results}
\subsection{Inclination of the Wavefront}\label{inclination}
The local morphology of the EUV wavefronts in the low corona on the solar limb is clearly shown in Figure \ref{stillMaps}.
Obvious inclination of the primary wavefronts is observed for five of the six events, except for the 2013 May 13 event.
Figure \ref{stillMaps} (a)--(e) illustrate a typical frame of the EUV wavefront for the five events, respectively.
The dashed-circles mark the altitudes from 1.05 to 1.25 {\rsun} with a step of 0.05 {\rsun},
where time-distance profiles in position angles (measured counterclockwise from the solar North) are extracted.
The time-distance profiles for each event are shown in panel (a) of Figures \ref{k20110912}--\ref{k20201129}.
If a wavefront is forward inclined, the wavefront at a larger altitude is ahead of that at a smaller altitude,
i.e., has a larger distance in the time-distance profile.
The time period during which the inclination is observed is indicated by the two vertical dotted lines in panel (a) of Figures \ref{k20110912}--\ref{k20201129}
(except Figure \ref{k20130513} because wavefront inclination is not obvious for the 2013 May 13 event).
For each time in the period, the EUV wavefront is determined by a linear fit of the points at the five altitudes from 1.05 to 1.25 {\rsun}.
For the 2013 November 7 and 2020 November 29 events, the point at 1.25 {\rsun} is excluded from the fit
because the position angle at 1.25 {\rsun} is very close to that at 1.20 {\rsun}
(see panel (a) of Figures \ref{k20131107} and \ref{k20201129}) indicating the wavefront above 1.20 {\rsun} is not notably inclined.
In this study, the inclination angle is defined as the acute angle
between the fitted wavefront and the tangent at the average position angle for each time.
The average inclination angle is calculated over the time period indicated
in panel (a) of Figures \ref{k20110912}--\ref{k20201129} (except Figure \ref{k20130513}).
The average inclination angles for the five events are given in Column (5) of Table \ref{tablist}
and are denoted in Figure \ref{stillMaps}(a)--(e).

There is a coronal hole at the north pole for the 2015 May 5 event, which transmits the EUV wave.
The EUV wavefront in the coronal hole can only be observed at lower altitudes of 1.05 and 1.10 {\rsun}.
As displayed in Figure \ref{k20150505}(a), at the two altitudes inside the coronal hole,
the distance between the two time-distance profiles becomes larger,
indicating that the wavefront is more inclined than the primary wavefront.
The inclination angle is estimated, although with only two points at the two altitudes, using a linear fit mentioned above.
Then we get the average inclination angle of the transmitted wavefront for the 2015 May 5 event,
which is $38\pm5$\degr and is given in Column (5) of Table \ref{tablist}.

For the famous 2017 September 10 event, the EUV wave is transmitted by the two polar coronal holes \citep{HuLZ2019ApJ},
and the southward wavefront is investigated.
Inside the south polar coronal hole, the wavefront is more obscured than outside and has a very high apparent speed
(see Figure \ref{k20170910} and Section \ref{kinematics}).
The position angles in the time-distance profiles of the wavefront have larger uncertainty,
and thus the wavefront inclination angle is not calculated inside the coronal hole.
However, after the EUV wave has left the coronal hole and has traveled to the other side of the Sun,
the wavefront becomes relatively clearer and has a lower apparent speed.
Figure \ref{stillMaps}(f) presents a running-difference image of the transmitted wavefront after it leaves the coronal hole.
The transmitted wavefront is detected at 3 altitudes from 1.05 to 1.15 {\rsun} (see Figure \ref{k20170910}(a)),
and the average inclination angle outside the coronal hole is estimated to be $38\pm1$\degr
(noted in Column (5) of Table \ref{tablist} and in Figure \ref{stillMaps}(f)).

For five of the six selected EUV wave events, wavefront inclination in the low corona is clearly observed,
and the inclination angle of coronal-hole transmitted wavefronts of two events is estimated for the first time.
As indicated in Column (5) of Table \ref{tablist},
the inclination angles of primary wavefronts for four events are around 60\degr{},
while the 2017 September 10 event has a larger inclination angle of $75\pm4$\degr{}.
The inclination angles of the coronal-hole transmitted wavefronts of the 2015 May 5 and 2017 September 10 events are both around 38\degr{}.
{We note that the 2017 September 10 event has the most intense flare,
but the flares of the 2013 November 7 and 2020 November 29 events are partially occulted
(as the eruption sites are slightly behind the limb).
So the actual flare energy of the events can hardly be compared.
Therefore, it is unclear whether the relatively larger inclination angle (i.e., being less inclined) of the primary wavefront
for the 2017 September 10 event is related to its observed strong flare intensity.}

{Assume that the 3D geometry of the EUV wavefront is generally symmetrical with respect to the solar meridian plane,
and thus the observed wavefront on the limb is approximately in the same meridian plane as the eruption site.
If the eruption site deviates by a small angle $\alpha$ from the meridian plane of the limb (i.e., the longitude of 90\degr{}),
the observed altitude of the wavefront at the solar equator in the EUV images
equals the actual altitude multiplied by $\cos \alpha$.
In this study, for all the selected events,
the angle $\alpha$ between the longitudes of the eruption site and the limb is below 10\degr{},
and the relative deviation of the actual altitude from the observed altitude is $1 - \cos \alpha \lesssim 1.5\%$.
This deviation decreases with the latitude,
which is even smaller for wavefronts far from the equator.
This tiny deviation applies equally to all altitudes,
and does not significantly affect the determination of the inclination angle and especially of its average.
The effect of the deviation on the estimate of the speed and its average
in the following Section \ref{kinematics} is also negligible.}

\subsection{Kinematics of the Wavefront}\label{kinematics}
The apparent speeds of the EUV wavefront at altitudes from 1.05 to 1.25 {\rsun} for each event
are from the derivative of the time-distance profiles.
The time-distance profiles in this study are from the visually determined leading edge of the wavefront.
The speed derived from visual investigation is consistent with that from a Gaussian fit of the wavefront intensity profile
\citep{DownsWL2021ApJ}.
The wavefront speeds ($v_{\text{w}}$) are plotted with solid curves in panels (b)--(f) of Figures \ref{k20110912}--\ref{k20201129},
which are then compared with the fast magnetosonic speed ($v_{\text{fms}}$) at each altitude.
The fast magnetosonic speed is obtained from the output of the Magnetohydrodynamic Algorithm outside a Sphere (MAS) 3D coronal MHD simulation
provided by the Predictive Science Inc.\footnote{The MHD output data are available from \url{https://www.predsci.com/mhdweb/data_access.php}.
The model option for the output used in this study is ``Thermodynamic with Heating Model 2''.
Descriptions of the model can be found at \url{https://www.predsci.com/corona/model_desc.html} (accessed 2024 September 9).}.
The fast magnetosonic Mach number, $M_{\text{fms}} = v_{\text{w}} / v_{\text{fms}}$, is calculated and plotted with dashed curves
in panels (b)--(f) of Figures \ref{k20110912}--\ref{k20201129}.
The speeds $v_{\text{w}}$ are averaged over the time span
specified by the horizontal solid line in panel (a) of Figures \ref{k20110912}--\ref{k20201129},
and then averaged over the five altitudes.
The twice averaged speeds are given in Column (6) of Table \ref{tablist}.
The details of the kinematics of the EUV wavefronts in the low corona for the selected six events are presented below.

For the 2011 September 22 event, as shown in Figure \ref{k20110912},
the speed at each altitude generally does not vary significantly with distance in position angle.
From the start to position angle $\sim$125\degr,
the wavefront is moving in a high fast-magnetosonic-speed region
(probably a small narrow coronal hole as seen in AIA 211 {\AA} images but not given here),
and the wavefront speed is below 400 \kmps{}.
Therefore, the Mach number is relatively low, below 1.
After leaving the high-speed region, the Mach number increases to above 1 for the wavefront at altitudes 1.05 -- 1.15 \rsun.
For the altitudes 1.20 and 1.25 \rsun, the wavefront again enters another high fast-magnetosonic-speed region
caused by the boundary of another coronal hole at the south pole (also visible in AIA 211 {\AA} images but not presented here).
This makes the Mach number at 1.20 and 1.25 \rsun{} below 1 throughout the propagation.
{As displayed in Figure \ref{k20110912}(a) and discussed in Section \ref{inclination},
between 10:44 and 10:52 UT, the wavefront is forward inclined.}
The actual speed is in the direction perpendicular to the wavefront,
which should be smaller than the apparent speed when the wavefront is forward inclined.
The actual speed can be estimated by the apparent speed multiplied by the sine of the inclination angle.
For the 2011 September 22 event, the inclination angle is $64\pm3$\degr,
which means that the actual speed could be $\sim10\%$ lower than the apparent speed when the wavefront is forward inclined ($\sin(64\degr{})\approx 0.9$).
The actual speed and its corresponding Mach number are plotted with cyan curves in panels (b)--(f) of Figure \ref{k20110912}.

For the 2013 May 13 event, the EUV wave starts at a high speed above 600 \kmps{} at all altitudes,
which can be seen in Figure \ref{k20130513}.
At altitudes 1.05 and 1.1 \rsun,
the speed rapidly decreases to $\sim$300 \kmps{} around 02:18 UT near position angle 35\degr{}.
Then the speed starts to increase near position angle 30\degr{} and reaches above 400 \kmps{} around position angle 15\degr{},
which may be caused by local loop systems at lower altitudes
\citep[similar to][]{DaiDC2012ApJ}.
{A group of low-corona loops near position angle 20\degr{} are obtained using a potential-field source-surface (PFSS) model
and are roughly consistent with a coronal cavity in AIA 211 \AA{}.
These are presented in Figure \ref{loops}(a)--(b) in Appendix \ref{pfss}.}
After passing position angle 20\degr{}, the wavefront is only visible at altitudes 1.05 and 1.10 \rsun{}.
Near position angle 0\degr{} around 02:36 UT,
the wavefront at the two altitudes decelerates to below 200 \kmps{} and eventually disappears.
This is possibly due to a coronal hole at the north pole, which can halt the propagation of an EUV wave
\citep[e.g.,][]{ThompsonPG1998GeoRL}.
The polar coronal hole cannot be identified in the AIA images,
but it is consistent with the increased fast magnetosonic speed given by the MHD model
(see the green dash-dotted curves in Figure \ref{k20130513}).
Because of the low fast magnetosonic speed, the Mach numbers at all altitudes are above 2
before the EUV wave reaches the coronal hole.
No clear wavefront inclination is observed for this event.

For the 2013 November 7 event, the speeds have a decreasing trend
and show larger fluctuations at lower altitudes (see Figure \ref{k20131107}).
The bump in the speed profiles near position angle 215\degr{}
could possibly again be induced by a local loop system.
The profile of fast magnetosonic speed there also has a small bulge,
which is possibly the indication of the local loop system.
{A set of loops in the low corona on the limb are derived by the PFSS model,
which are possibly associated with two prominences in AIA 304 {\AA}.
These are displayed and detailed in Figure \ref{loops}(c)--(d) in Appendix \ref{pfss}.}
After the bump in the speed profile, the speed at all altitudes decreases gradually
to below the fast magnetosonic speed near position angle 190\degr{}.
The green dash-dotted curve in Figure \ref{k20131107} shows that
the EUV wavefront is leaving its active region of a relatively high fast magnetosonic speed.
During most of the travel, the Mach number is above 1.5, which means that the wavefront for this event is a shock.
Considering the inclination angle of $65\pm2$\degr,
the actual speed is also $\sim10\%$ lower than the apparent speed in the time span
specified by the two vertical dotted lines in Figure \ref{k20131107}(a).

For the 2015 May 5 event, the speed profiles are shown in Figure \ref{k20150505},
showing a notable deceleration in the initial stage.
Near position angle 10\degr{} close to the north pole,
the EUV wavefront encounters a coronal hole and the EUV wave is transmitted by the coronal hole.
However, the fast magnetosonic speed inside the coronal hole near the north pole
(for position angles smaller than 10\degr{})
derived from the MHD model is not as high as expected for a coronal hole.
This could be due to the large measurement uncertainty of magnetic fields in the polar region.
A rapid deceleration of the wavefront at altitudes 1.20 and 1.25 \rsun{} is observed,
which also indicates the existence of the coronal hole.
As shown in Figure \ref{k20150505}(e)--(f),
the deceleration occurs a little earlier before the wavefront reaches position angle 10\degr{}.
It is possibly because the magnetic fields of a coronal hole are diverging,
which induces a larger angular width of the coronal hole at higher altitudes
\citep{AltschulerTO1972SoPh,KoppH1976SoPh}.
The transmitted wavefront in the coronal hole is discernible only at 1.05 and 1.10 \rsun{},
which has a significantly increased speed.
The inclination angle of $38\pm5$\degr{} is much smaller than that of the primary wavefront
($66\pm1$\degr{}, see Column (5) of Table \ref{tablist}).
This means that the actual speed of the wavefront in the coronal hole can be $\sim40\%$ lower than the apparent speed.
As indicated by the cyan curves in Figure \ref{k20150505}(b)--(c),
the actual speed calculated by taking into account the inclination
is comparable to the speed of the primary wavefront before entering the coronal hole.
The primary wavefront is also forward inclined,
and the actual speed can also be $\sim10\%$ lower than the apparent speed.
Before entering the coronal hole, the Mach number is mostly above 1.5 at all altitudes.
However, the Mach number inside the coronal hole is not given in Figure \ref{k20150505}
because of the inconsistent fast magnetosonic speed from the MAS MHD model.

For the notable 2017 September 10 global event,
the transmission of the EUV wavefront by two polar coronal holes has been reported in previous researches
\citep[e.g.,][]{LiuJD2018ApJ,HuLZ2019ApJ}.
Before entering the south polar coronal hole,
the wavefront speed increases with the altitude from $\sim$600 \kmps{} to $\sim$1000 \kmps{}
(see Figure \ref{k20170910}(b)--(f)).
From the inbound boundary of the coronal hole (near position angle 210\degr{}),
only the wavefront at altitudes of 1.05--1.15 \rsun{} is visible.
Inside the coronal hole, the apparent wavefront speed is very high,
and ranges from $\sim$1000 \kmps{} to even $\sim$3000 \kmps{},
which is comparable to the estimate of \citet{LiuJD2018ApJ}.
After leaving the south polar coronal hole,
the wavefront undergoes a gradual deceleration before it collides with the wavefront
transmitted by the north polar coronal hole \citep[see][]{HuLZ2019ApJ}.
Behind the outbound boundary of the south polar coronal hole (near position angle 140\degr{}),
the apparent wavefront speed is still $\sim700$ \kmps{} and the Mach number can be nearly 3 (see Figure \ref{k20170910}(b)--(d)).
Considering the inclination angle of $38\pm1$\degr{} for the transmitted wavefront,
the actual speed can be below 500 \kmps{} and the Mach number can be below 2
(see the cyan curves in Figure \ref{k20170910}(b)--(d)).
Inside the coronal hole the inclination angle is not estimated
because the wavefront is too diffuse to obtain a reliable angle with visible inspection.
If we assume the inclination angle is the same as the transmitted wavefront behind the coronal hole,
the actual wavefront speed inside the coronal hole can also be $\sim40\%$ lower than the apparent speed.
In the initial stage, the inclination angle of the primary wavefront is $75\pm4$\degr{},
which means the wavefront is less inclined than those of other events
and the actual speed is approximately equal to the apparent speed.

For the 2020 November 29 event, the wavefront speed profiles are displayed in Figure \ref{k20201129}.
There is a bump in the speed profiles at 1.05 and 1.10 \rsun{} near position angle 65\degr{},
which spatially corresponds to the increase of fast magnetosonic speed at the lower altitudes
(see the dash-dotted curves in Figure \ref{k20201129}(b)--(c)).
Similar to the speed profile bumps in the 2013 May 13 and 2013 November 7 events,
the speed increase is also possibly attributed to a local loop system.
{The local loop system is possibly associated with a coronal bright point with bipolar magnetic fields near latitude 25\degr{},
which is illustrated in Figure \ref{loops}(e)--(f) in Appendix \ref{pfss}.
Near position angle 40\degr{}, the EUV wavefront at altitudes above 1.1 \rsun{} vanishes,
and the wavefront speed at 1.05 \rsun{} is slightly elevated
along with the MHD-derived fast magnetosonic speed (see Figure \ref{k20201129}(b)).
As shown in Figure \ref{loops}(e)--(f) in Appendix \ref{pfss},
a small filament is observed near position angle 35\degr{} in SDO/AIA 211 {\AA} images,
indicating another loop system that is similar to the coronal cavity with a high characteristic speed in
\citet{LiuON2012ApJ}.
This loop system may be the reason that the wavefront speed is elevated at 1.05 \rsun{} after passing position angle 35\degr{}.}
The Mach number varies from 1.5 to 2.5 during most of the propagation.
In the initial stage, the average inclination angle of the primary wavefront is $61\pm1$\degr{},
which indicates that the actual speed is $\sim10\%$ lower than the apparent speed,
like the 2011 September 22, 2013 November 7, and 2015 May 5 events.

\begin{figure}[ht!]
\centering
\includegraphics[width=0.47\textwidth]{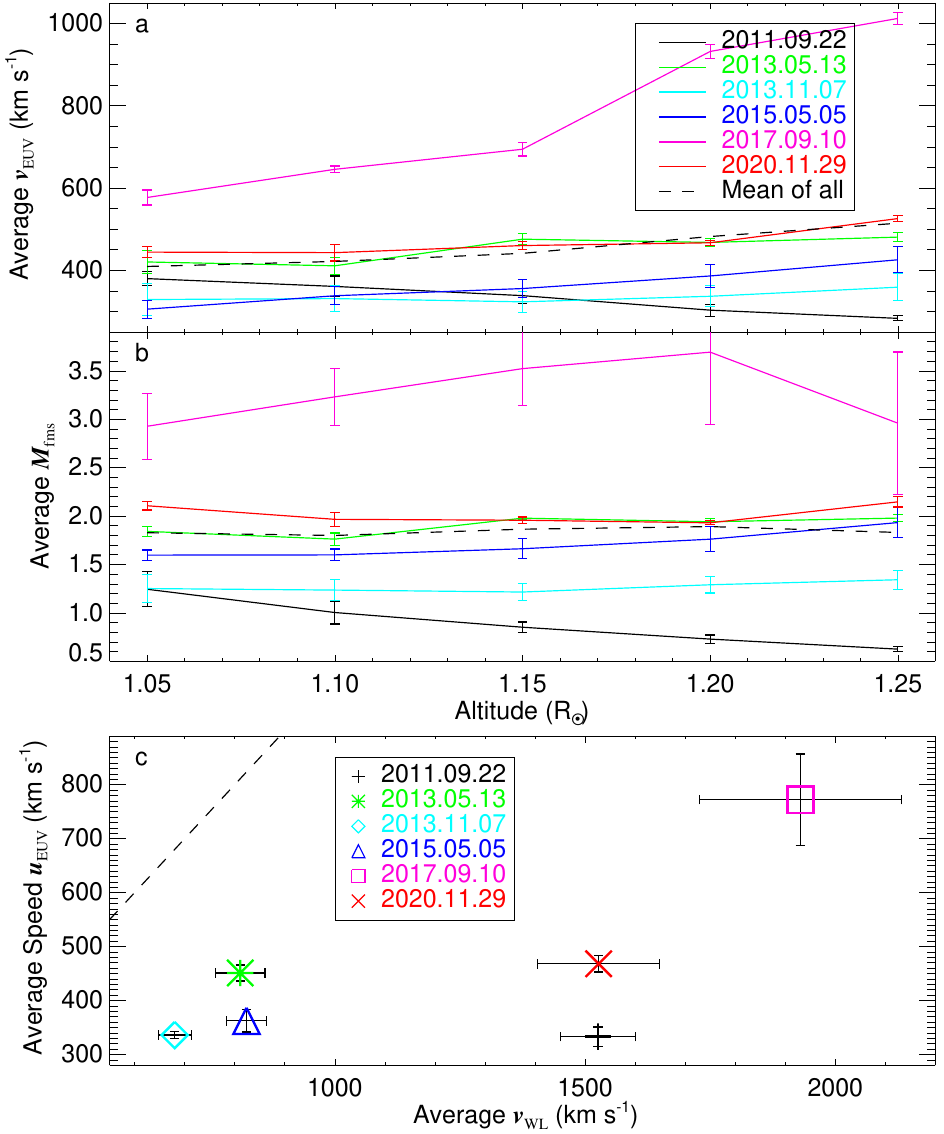}
\caption{\label{vles}{(a) and (b) Average speed ($v_{\text{EUV}}$) and average fast magnetosonic Mach number ($M_{\text{fms}}$)
of the EUV wavefronts at altitudes from 1.05 to 1.25 {\rsun};
the dashed curve represents the mean value of all the events at these altitudes;
the average time span is marked by the horizontal solid line in panel (a) of Figure \ref{k20110912}--\ref{k20201129} for each event, respectively.
(c) The average speed ($u_{\text{EUV}}$) of the EUV wavefronts compared with the lateral expansion speed of white-light CME-driven shocks ($v_{\text{WL}}$);
$u_{\text{EUV}}$ is the averaged the speeds in panel (a) over all the five altitudes for each event,
which is also listed in Column (6) of Table \ref{tablist};
the average lateral-expansion speed $v_{\text{WL}}$ is from the fit of the ellipsoid model (see the text)
and is given in Column (7) of Table \ref{tablist};
the dashed line is a one-to-one visual guide.}}
\end{figure}

The speed of the EUV wavefront is averaged at each altitude for all the events,
over the time span marked with the horizontal solid line in panel (a) of Figures \ref{k20110912}--\ref{k20201129}.
{The averaging time span is determined by the requirement that the wavefront must be observed at all five altitudes.}
In the averaging, the apparent speeds of the inclined wavefronts
are replaced with the actual speeds by removing the effect of inclination.
The average speeds $v_{\text{EUV}}$ of the six events are shown in Figure \ref{vles}(a),
where the dashed curve represents the mean value of all the events.
The average speeds of five events are in the range of 300--500 \kmps{},
but for the 2017 September 10 event the average speed increases with the altitude from $\sim600$ to $\sim1000$ \kmps{}.
The large average speed for the 2017 September 10 event
indicates the extremely rapid lateral expansion of the CME reported in previous studies
\citep[e.g.,][]{LiuJD2018ApJ,VeronigPD2018ApJ}.

The fast magnetosonic Mach number ($M_{\text{fms}}$) of the wavefronts for the six events
is also averaged in the same way as the speed,
and is plotted in Figure \ref{vles}(b).
The average $M_{\text{fms}}$ for all events is $\sim1.7$ as indicated by the dashed curve.
For the 2017 September 10 event, $M_{\text{fms}} \approx 3$ is outstandingly large at all the altitudes,
which is in accordance with the high speed of the wavefront.
For the 2011 September 22 event, the average $M_{\text{fms}}$, especially at higher altitudes, is very low and below 1,
which is because the wavefront is traveling in a high fast-magnetosonic-speed region (see the dash-dotted curves in Figure \ref{k20110912}(b)--(f)).
For the other four events, the average $M_{\text{fms}}$ is in a range of $\sim1.2$--$\sim2.1$.
Generally the average $v_{\text{EUV}}$ and $M_{\text{fms}}$ of the four events
do not vary significantly with the altitude in the low corona from 1.05 to 1.25 \rsun.

The speed $v_{\text{EUV}}$ is then averaged over all the altitudes for each event,
and the average is noted as $u_{\text{EUV}}$ in Figure \ref{vles}(c).
In Figure \ref{vles}(c) $v_{\text{WL}}$ is the average speed of the lateral expansion of the white-light CME-driven shock.
The lateral expansion speed is fitted with an ellipsoid model developed by \citet{KwonZO2014ApJ}.
This ellipsoid model has seven free parameters:
the height, longitude, and latitude of the ellipsoid center;
the three semi-principal axes; and the rotation angle of the ellipsoid.
More details of the model can be found in \citet{KwonZO2014ApJ}.
For the 2017 September 10 and 2020 November 29 events,
the lateral expansion speeds of the CME-driven shocks are adopted from the results using the same model in
\citet{LiuZZ2019ApJ} and \citet{ChenLZ2022ApJ}, respectively.
For the other four events,
the two semi-principal axes perpendicular to the radial direction are set to equal
and the longitude is set to that of the solar limb,
which reduces three free parameters (one semi-principal axis, the rotation angle, and the longitude).
In this condition, the SOHO/LASCO observation from one viewpoint is enough to fit the shock speed with the ellipsoid model.
The lateral expansion speed of the white-light shock
is measured at the vertex of the principal axis perpendicular to the radial direction in the ellipsoid model.
The average speed of the shock ($v_{\text{WL}}$)
is compared with the average speed of the EUV wavefront ($u_{\text{EUV}}$) in Figure \ref{vles}(c),
and no obvious correlation is found between the two average speeds.
{Because the CME and its driven shock appear in LASCO C2 in the later stage of the EUV wavefront propagation,
there are no more than two C2 images in the time span during which the average speed of the EUV wavefront is calculated.
Therefore, the average speeds of the EUV wavefront ($u_{\text{EUV}}$)
and the white-light shock ($v_{\text{WL}}$) are not from the same time interval.
This may be part of the reason why the two average speeds are not well correlated,
as will be discussed in Section \ref{disSpeed}.}

\subsection{Coupling of the CME Boundary and the EUV Wave}\label{coupling}

\begin{figure*}[ht!]
  \centering
  \includegraphics[width=\textwidth]{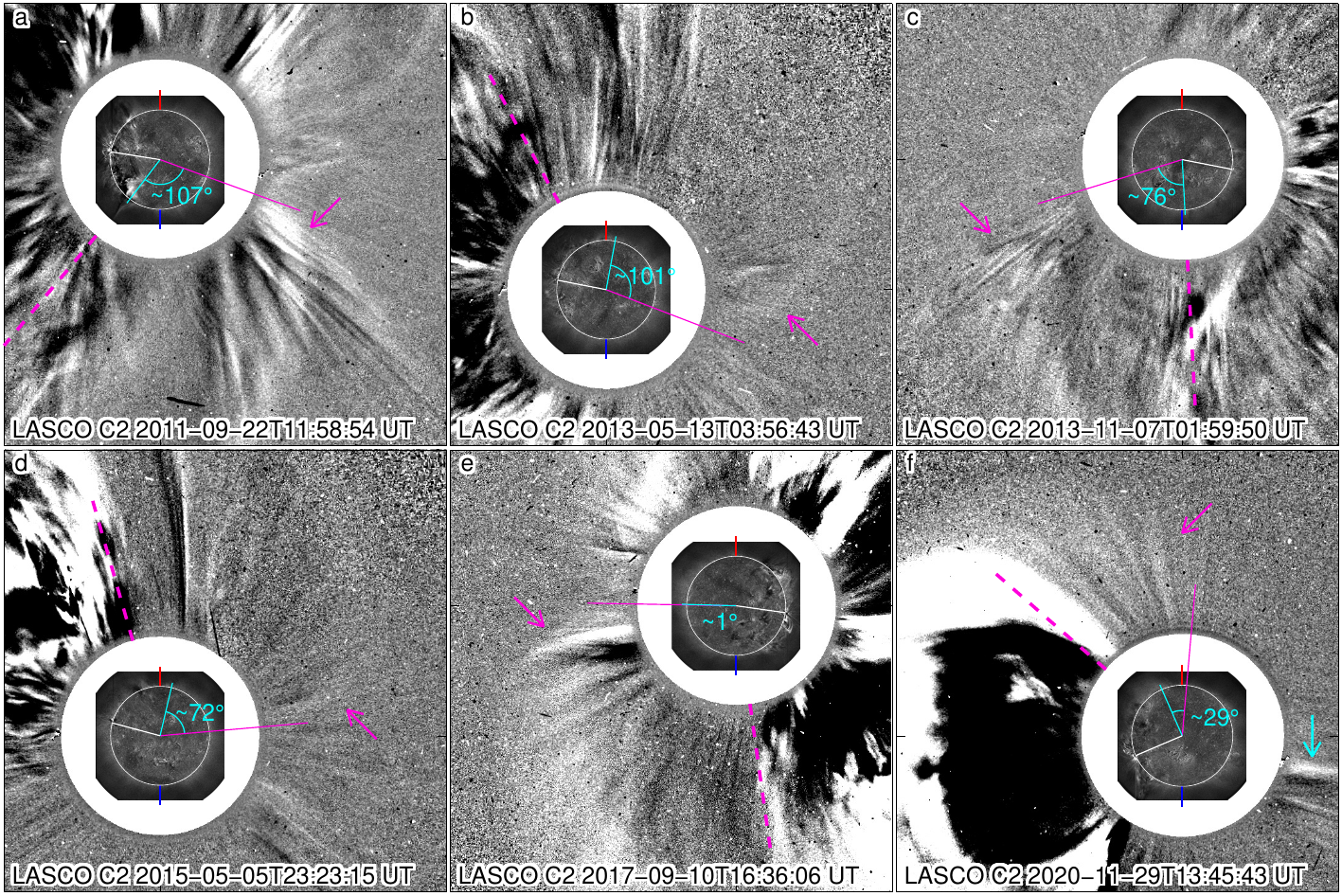} 
  \caption{\label{c2aia}{Angular distances of the EUV wavefront, CME boundary, and white-light shock flank for the six events.
  Each panel contains a composite of running-difference images of SOHO/LASCO C2 and SDO/AIA 211 {\AA}.
  The C2 time is overlaid on the image, and the AIA time is around when the EUV wave disappears.
  The white line marks the position angle of the eruption site, which is given in Column (2) of Table \ref{tabrange}.
  The cyan line indicates the position angle where the EUV wavefront vanishes at all altitudes,
  from which the maximum travel distance of the EUV wavefront is obtained (see Column (3) of Table \ref{tabrange}).
  The thick pink dashed line denotes the position angle of the CME boundary at the inner edge of the field of view (FOV) of LASCO C2
  around the time when the EUV wave disappears,
  and the corresponding angular distance to the eruption site is given in Column (4) of Table \ref{tabrange}.
  The pink solid line marks the position angle of the white-light shock flank at the inner edge of LASCO C2 FOV
  when the shock flank reaches its maximum distance from the eruption site (also provided in Column (5) of Table \ref{tabrange}).
  The cyan arc and the corresponding text indicate the angular distance between the cyan line and the pink solid line.
  The pink arrow points to the white-light shock flank,
  which shows that the shock has laterally expanded to the opposite side with a maximum distance larger than 90\degr{}.
  The cyan arrow in panel (f) indicates the southward shock flank interacting with a streamer.}}
\end{figure*}

\begin{deluxetable*}{ccccccc}
  \tablenum{2}
  \tablecaption{Travel distance of EUV wave, CME boundary, and white-light shock flank \label{tabrange}}
  \tablewidth{0pt}
  \tablehead{
  \colhead{Event Number} & \colhead{Site} & \colhead{EUV} & \colhead{CME} & \colhead{WL} & \colhead{EUV-CME} & \colhead{EUV-WL}  \\
  \colhead{(YYYY-MM-DD)} & \multicolumn{6}{c}{(All units are in degrees of position angle)}}
  \colnumbers
  \startdata
  2011-09-22 & 81   & 62  (143) & 59 (140) & 169 (250) & 3   & $-$107 \\
  2013-05-13 & 79   & 89  (350) & 50 (29)  & 190 (249) & 39  & $-$101 \\
  2013-11-07 & 259  & 76  (183) & 76 (183) & 152 (107) & 0   & $-$76  \\
  2015-05-05 & 75   & 88  (347) & 59 (16)  & 160 (275) & 29  & $-$72  \\
  2017-09-10 & 262  & 174 (88)  & 74 (188) & 173 (89)  & 100 & 1      \\
  2020-11-29 & 113  & 89  (24)  & 64 (49)  & 118 ($-$5)& 25  & $-$29  \\
  \enddata
  \tablecomments{Column (2) is the position angle of the eruption site.
  Column (3) gives the maximum angular travel distance of the EUV wavefront from the eruption site in position angle.
  Column (4) provides the angular distance of the CME boundary at the inner edge of SOHO/LASCO C2 field of view around when the EUV wave disappears.
  Column (5) is the maximum angular distance of the white-light shock flank observed by LASCO C2.
  Numbers in the parentheses of Columns (3)--(5) are the position angles corresponding to the distances,
  which are also marked with cyan, thick pink dashed, and pink solid lines in Figure \ref{c2aia}, respectively.
  Column (6) is from Columns (3) minus (4), and Column (7) is from Columns (3) minus (5).
  The position angles of the CME boundary and the shock flank
  are determined by visual inspection and may have an uncertainty of $3\degr$.}
\end{deluxetable*}

In the combined interpretation of an EUV wave,
the wave component decouples from the CME driver (i.e., the non-wave component)
before the EUV wave fades out,
after which the wave component propagates freely.
The separation of the EUV wavefront and the CME boundary,
demonstrating the decoupling, has been observed in a plethora of events
\citep[e.g.,][]{PatsourakosV2009ApJ,ChengZO2012ApJ,DaiDC2012ApJ,LiberatoreLV2023ApJ}.
This means that the angular distance between the two components should be notable,
when they are observed on the solar limb.
The position angle where the EUV wavefront disappears at all altitudes from 1.05 to 1.25 \rsun{}
is marked by the cyan line in Figure \ref{c2aia}.
The maximum travel angular distance of the EUV wavefront from the eruption site is estimated
and given in Column (3) of Table \ref{tabrange}.
The position angle of the CME boundary corresponding to the EUV wavefront
is also estimated in the running-difference SOHO/LASCO C2 image
around the time when the wavefront disappears,
which is marked by the thick pink dashed line in Figure \ref{c2aia}.
The angular distance from the eruption site to the CME boundary is presented in Column (4) of Table \ref{tabrange}.
The difference between the two distances,
which is equal to the angle between the cyan line and the thick pink dashed line in Figure \ref{c2aia},
is given in Column (6) of Table \ref{tabrange}.

For the 2011 September 22 and 2013 November 7 events,
we can see the angular distance between the EUV wavefront and the CME boundary is close to zero (see Column (6) of Table \ref{tabrange}),
which suggests that the two features are coupled until the EUV wave fades out.
For the other four events, the angular distance between the EUV wavefront and the CME boundary is significantly greater than 0\degr{},
which will be discussed separately below.

For the 2013 May 13 event, the CME boundary is around position angle 29\degr{}
(see Column (4) of Table \ref{tabrange}) and is 39\degr{} behind the EUV wavefront when the wavefront disappears.
As discussed in Section \ref{kinematics} and shown in Figure \ref{k20130513},
around position angle 30\degr{} the wavefront speed increases possibly because of local loop systems in the lower corona,
from which the EUV wavefront decouples from the CME driver.
This is similar to the case in \citet{DaiDC2012ApJ} where the CME boundary slows down and decouples from the EUV wavefront at a loop system.

For the 2015 May 5 event, the CME boundary is near position angle 16\degr{} and is 29\degr{} behind the EUV wavefront
when the wavefront at 1.05 \rsun{} fades out.
Around position angle 10\degr{} there is a coronal hole which can halt the expansion of the CME
\citep[e.g.,][]{ChenPF2009ApJ} but transmits the EUV wave in the low corona (see Figure \ref{k20150505}).
At the higher altitudes (1.15--1.25 \rsun{}), the wavefront decelerates before it reaches position angle 10\degr{},
which could be because the coronal hole has diverging magnetic fields and has a larger angular width at the higher altitudes
\citep{AltschulerTO1972SoPh,KoppH1976SoPh}.
Before the EUV wavefront encounters the coronal hole, the Mach number is above 1.5 and even above 2,
which suggests that the wavefront is not free and is possibly continuously driven by the expanding CME.
The CME boundary loops cannot enter the coronal hole,
and thus the EUV wavefront decouples from the CME and propagates freely inside the coronal hole.
As denoted in Figure \ref{c2aia}(d) and Column (6) of Table \ref{tabrange},
the CME boundary is 29\degr{} behind when the EUV wavefront vanishes.

For the 2017 September 10 event,
the CME boundary is first observed near the south polar coronal hole around 16:12 UT
when the EUV wavefront has already been transmitted to the other side of the coronal hole (see Figure \ref{k20170910}).
The EUV wavefront travels a short distance before reaching the coronal hole,
and has a large Mach number above 3.
It is in the early stage of the CME expansion,
where the fast expanding CME is driving and coupled with the EUV wavefront.
The CME boundary is halted at the inbound boundary of the coronal hole
while the decoupled EUV wavefront is transmitted to the opposite side of the Sun.
This results in a 100\degr{} large angular distance between them
(see Figure \ref{c2aia}(e) and Column (6) of Table \ref{tabrange}).

For the 2020 November 29 event,
as discussed in Section \ref{kinematics} and shown in Figure \ref{k20201129},
after the EUV wavefront reaches position angle 40\degr{},
the wavefront at altitudes above 1.05 \rsun{} vanishes.
Only the wavefront at 1.05 \rsun{} reaches position angle 24\degr{},
which is 25\degr{} ahead of the CME boundary (see Figure \ref{c2aia}(f) and Column (6) of Table \ref{tabrange}).
Due to a data gap,
the CME is first observed by SOHO/LASCO C2 around 13:23 UT
with its boundary $\sim$10\degr{} behind the EUV wavefront,
when the wavefront has faded out above 1.1 \rsun{} but is still visible at 1.05 \rsun{}.
This suggests that the CME decouples from the EUV wavefront before the wavefront fades out at all altitudes.

Due to the large cadence of the SOHO/LASCO C2 images,
it is not possible to track the CME boundary during the whole propagation of the EUV wavefront.
For the 2011 September 22 and 2013 November 7 events,
the angular distance between the EUV wavefront and the CME boundary is approximately zero,
when the wavefront fades out at all altitudes.
Although the white-light CME structure below 2 \rsun{} cannot be observed by SOHO/LASCO C2,
the CME boundary is almost radial as illustrated in Figure \ref{c2aia}
and can extend straight downward to below 2 \rsun{}.
This is consistent with that the legs or the overlying loops of the CME boundary are stretched to straight
during the rising and expansion of a CME structure.
It is indicated that the EUV wavefront is still coupled with the CME boundary when the wavefront vanishes.
This suggests during the whole propagation,
the EUV wavefront is not propagating freely but is continuously driven by the expanding CME in the two events.

\subsection{Connection between the EUV Wavefront and the CME-driven Shock}\label{connection}
A CME-driven shock is a 3D structure that can be observed in white light in the high corona,
which connects to the EUV wavefront in the low corona after decoupling from the CME driver
\citep[e.g.,][]{VeronigMK2010ApJ,ChengZO2012ApJ,KwonZO2014ApJ,ZhuLK2018ApJ}.
Attributed to the transmission of coronal holes,
the EUV wavefront for the 2017 September 10 event
can travel simultaneously with the white-light shock on the solar limb to the other side of the Sun,
during which the white-light shock is ``incurvated'' to the low corona
and continuously connected with the EUV wavefront \citep{HuLZ2019ApJ}.
This connection is also indicated by the cyan and pink solid lines in Figure \ref{c2aia}(e),
which shows that the angular distance between the EUV wavefront and the shock flank is,
$\sim$1\degr{}, close to zero (see Column (7) of Table \ref{tabrange}).
However, for the 2011 September 22, 2013 May 13, 2013 November 7, and 2015 May 5 events,
the shock continues to expand laterally after the EUV wavefront fades out at all altitudes,
which results in a notable angular distance larger than 70\degr{} between the two features.
As indicated in Figure \ref{c2aia}(a)--(d) and Column (7) of Table \ref{tabrange},
the white-light shock flank is $\sim$107\degr{}, $\sim$101\degr{}, $\sim$76\degr{}, and $\sim$72\degr{} ahead of the EUV wavefront
for the above four events, respectively.
For the 2020 November 29 event, the EUV wavefront and the CME-driven shock are both asymmetric.
The northward shock flank, corresponding to the analyzed EUV wavefront, is faint,
and its angular position is roughly estimated to be only $\sim$29\degr{} ahead of the EUV wavefront.
However, as indicated by the cyan arrow in Figure \ref{c2aia}(f), the southward shock flank has traveled a large distance,
which could be more than 90\degr{} from the final position angle of the southward EUV wavefront.
Because the southward EUV wavefront for this event is visible for only a short range,
it is not selected for analysis here.

The white-light CME-driven shocks in the corona for all the events have expanded laterally
to the side opposite the eruption site on the solar limb.
For the unique 2017 September 10 event,
the EUV wave is transmitted by the polar coronal holes and is connected with the white-light shock on the opposite side
\citep{HuLZ2019ApJ}.
For the other five events,
the EUV wavefront travels for a distance no larger than 90\degr{} (see Column (3) of Table \ref{tabrange}),
and the white-light shock flank is still traveling after the EUV wavefront has vanished at all altitudes.
As noted in Column (7) of Table \ref{tabrange},
the white-light shock flank in the later stage is more than 70\degr{} away from the final position angle of the EUV wavefront
(for the 2020 November 29 event this is true for the southward shock flank as discussed in the previous paragraph).
The position angle of the shock flank is estimated at $\sim$2 \rsun{} in the SOHO/LASCO C2 running-difference images,
which cannot be geometrically linked by a straight line to the EUV wavefront at $\sim$1 \rsun{}.
This implies that the EUV wavefront is unlikely connected to the white-light shock flank for the five events,
and is no longer the footprint of the shock after the shock has propagated to the opposite side.

\section{Discussions}\label{discussions}
Based on the imaging observations of SDO/AIA 211 {\AA} and SOHO/LASCO C2,
we have investigated the EUV wavefronts in the low corona for six selected EUV wave events.
The inclination, speed, coupling with the CME, and spatial connection with the shock have been analyzed,
which are summarized and discussed below in this section.

\subsection{Inclination of the Primary and Transmitted Wavefronts}\label{disInclination}
The inclination of the primary EUV wavefront is evident on the solar limb
in the initial propagation stage for five of the six events (see Figure \ref{stillMaps}).
The wavefront inclination is considered a natural result
of the increasing fast magnetosonic speed with the altitude in the low corona,
which is applicable to freely propagating EUV waves
\citep[e.g.,][]{UchidaY1968SoPh,WarmuthVM2004AA,AfanasyevU2011SoPh,KwonV2017ApJ,MannV2023AA}.
In our events, the inclination of the EUV wavefront occurs in the early stage,
in which the CME structure is rapidly expanding and is still coupled with the EUV wavefront.
This means that the inclination of the EUV wavefront illustrates the forward inclined loops
at the boundary of the expanding CME.
If the inclination is caused by the fast magnetosonic speed increasing with the altitude,
the inclination angle, defined in this study, should decrease with distance according to models
\citep[e.g.,][]{UchidaY1968SoPh,AfanasyevU2011SoPh,KwonV2017ApJ}.
This requires that the separation between the time-distance profiles,
as displayed in panel (a) of Figure \ref{k20110912}, \ref{k20131107}--\ref{k20201129},
should also increase with time.
However, except for the 2017 September 10 event, this increase is not seen in the time-distance profiles.
For the 2017 September 10 event, the inclination is only observed briefly in the early stage of the propagation,
when the CME is still rapidly expanding and is coupled with the EUV wavefront.
The inclination angle of the primary wavefront for the five events ranges in $61\degr$--$75\degr$ (see Column (5) of Table \ref{tablist}),
which reflects the geometry of the forward inclined loops at the boundary of the expanding CME.
{The geometry may be related to a few factors, such as
the initial magnetic configuration of the erupted flux rope and the source region;
the eruption direction when the CME has a non-radial motion in the corona
\citep[e.g.,][]{MoestlRF2015NatCo,WangLD2015ApJ,HuLW2017ApJ,WangHL2020JGRA},
for which the loops on the propagation path may be more inclined toward the solar surface;
and the height of the magnetic reconnection that triggers the eruption
(i.e., near the eruption site the CME boundary loops expanded by a higher eruption might be more inclined).}

After being transmitted by a coronal hole,
the wavefront becomes more forward-inclined with the average inclination angle reduced to $38\degr$,
which is observed for the 2015 May 5 and 2017 September 10 events.
After entering the coronal hole, the EUV wavefront is visible
only at lower altitudes below 1.15 \rsun{} (see panel (a) of Figures \ref{k20150505}--\ref{k20170910}).
{The wavefront has already decoupled from the CME driver and becomes freely propagating
because a CME structure cannot enter a coronal hole
\citep[e.g.,][]{GopalswamyMX2009JGRA,LiuZH2019ApJS,SahadeCK2020ApJ}.}

The inclination angle behind the coronal-hole boundary is smaller than that of the primary wavefront coupled with the CME driver,
which can be attributed to the increase of the fast magnetosonic speed with the altitude in the low corona.
The inclination of a freely propagating EUV wavefront has also been observed in previous studies
\citep[e.g.,][]{LiuON2012ApJ,HouTW2022ApJ},
where the inclination angle (above 50\degr{}) is larger than that of the transmitted EUV wavefront in our study ($\sim38$\degr{}).
In our study the travel distance of the inclined transmitted wavefront is larger than 700 Mm,
which is much larger than those ($<$300 Mm) in \citet{LiuON2012ApJ} and \citet{HouTW2022ApJ}.
The smaller inclination angle could be attributed to that the angle decreases with the travel distance
as a result of the fast magnetosonic speed increasing with the altitude in the low corona.
This is also possibly because the fast magnetosonic speed inside the coronal hole
increases more rapidly with the altitude in the low corona than outside the coronal hole
(see the $v_{\text{fms}}$ increasing with altitudes in the coronal hole in Figure \ref{k20170910}),
which significantly increases the refraction and enlarges the inclination of the EUV wavefront.

The inclination of the EUV wavefront can significantly affect the measurement of the actual wavefront speed.
The speed at an altitude is measured from the apparent motion of the wavefront,
and the actual speed direction is perpendicular to the local wavefront.
If the wavefront is forward inclined, the propagation direction is toward the solar surface.
Thus, the actual speed is the apparent speed multiplied by the sine of the inclination angle,
where the inclination angle is defined in Section \ref{inclination} and is schematized in Figure \ref{stillMaps}.
For the primary wavefront as discussed above, the inclination angle is in the range of $61\degr$--$75\degr$,
whose effect on the estimate of the actual speed is around 10\% and can be negligible.
However, for the wavefront transmitted by coronal holes in the 2015 May 5 and 2017 September 10 events,
the inclination angle is as small as $38\degr$ and the actual speed can be $\sim40\%$ lower than the apparent speed.
The effect of the inclination of the transmitted wavefront
on the estimate of the actual speed and Mach number is considerable.
For the 2015 May 5 event as shown in Figure \ref{k20150505}(b)--(c),
the wavefront speed inside the coronal hole multiplied by the sine of the inclination angle
becomes comparable to that outside the coronal hole.
For the 2017 September 10 event as shown in Figure \ref{k20170910}(b)--(d),
considering the inclination angle the Mach number of the transmitted wavefront is modified from $\sim$3 to below 2,
which is more reasonable for a freely propagating EUV wavefront behind a coronal hole.

\subsection{The Speed of the EUV Wavefront and its Comparison with the Shock Lateral Expansion Speed}\label{disSpeed}
With observations on the limb and excluding the projection effect,
it is observed that the kinematics of EUV wavefronts is not uniform at different altitudes.
For the 2013 May 13, 2013 November 7, and 2020 November 29 events,
the speeds at the low altitudes are locally elevated and are different from those at the higher altitudes,
which is possibly caused by local loop systems (see Section \ref{kinematics} {and Appendix \ref{pfss}).
Besides the elevation of wavefront speed by loops
\citep[e.g.,][]{DaiDC2012ApJ,LiuON2012ApJ,HuLZ2019ApJ},
an EUV wave can also manifest a diversity of phenomena when it encounters different coronal structures,
for example, reflection by coronal holes
\citep[e.g.,][]{GopalswamyYT2009ApJ,LiZY2012ApJ,ShenL2012ApJ,ZhouST2022AA},
refraction or transmission by active regions or coronal holes
\citep[e.g.,][]{VeronigTV2006ApJ,OlmedoVZ2012ApJ,ShenLS2013ApJ,LiuJD2018ApJ,LiuWL2019ApJ},
and interference after being transmitted by a coronal hole with a proper shape
\citep[e.g.,][]{ZhouSL2022ApJ,ZhouSY2024NatCo}.}
However, it was not noticed before that the speed is altered locally at only lower altitudes deviating from the speed at the higher altitudes,
which may induce distortion in the EUV wavefront.
If an EUV wavefront is crossing loop systems on the solar disk and is observed from a single viewpoint,
it will be difficult to obtain the actual speed and to distinguish the wavefront at different altitudes.
Additionally, the fast magnetosonic speed increases with the altitude in the low corona,
which leads to higher speeds at higher altitudes as well as the wavefront inclination \citep[e.g.,][]{WarmuthVM2004AA,AfanasyevU2011SoPh,DownsWL2021ApJ}.
These add more difficulty to estimating the actual speed of an on-disk EUV wavefront from single-viewpoint observations
and to using the speed in local coronal seismology.

At altitudes 1.05--1.25 \rsun{}, the average speeds of the primary EUV wavefronts are generally in the range of 300--500 \kmps{},
and their corresponding fast magnetosonic Mach numbers are in the range of 1.2--2.1 (see Figure \ref{vles}(a)--(b)),
where the exceptions are the 2017 September 10 and 2011 September 22 events.
For the 2011 September 22 event, because the EUV wavefront is traveling in a region of high fast magnetosonic speed,
the Mach number is below 1 at higher altitudes.
For the 2017 September 10 event, both the average speed and the Mach number are much higher than those of the other events,
which are also consistent with those derived from the MHD simulation of \citet{YangWF2021ApJ}.
Excluding the two exceptions, the Mach numbers are above 1 at altitudes 1.05--1.25 \rsun{},
which means that the EUV wavefront is a shock even at the coronal base from the initial stage
when the wavefront is coupled with the expanding CME.
The Mach number is calculated using the fast magnetosonic speed given by the MAS MHD simulation,
which is based on non-real-time data on the solar limb and could have larger uncertainty.
However, the average Mach numbers at different altitudes are still a useful parameter to indicate the shock formation.

The average speed of the lateral expansion of the CME-driven shock has been estimated using an ellipsoid model,
which shows no obvious correlation with the average speed of the EUV wavefront (see Figure \ref{vles}(c)).
The lateral expansion speed of the shock is measured at the vertex of the principal axis of the ellipsoid model,
which is perpendicular to the radial direction.
The measurement is based on the white-light images of the shock in the high corona,
when the primary EUV wavefront has already started to fade out
(although the EUV wave in the 2017 September 10 event is transmitted by coronal holes and propagates for a long time,
the primary EUV wave travels for only a short time during which the wavefront speed is calculated).
The EUV wavefront is an indicator of the early-stage expansion of the shock in the low corona,
which is coupled with the CME for the events in this study.
The EUV wavefront speed is measured in the early stage in the low corona
and the shock speed is measured in the later stage in the high corona,
which may explain the weak correlation between the two speeds.
This may be similar to the low correlation between the EUV wavefront speed and the CME radial speed
reported in previous studies \citep[e.g.,][]{NittaST2013ApJ,MuhrVK2014SoPh}.
In addition, inferred from the fact that the CME has a limit width and the shock has propagated to the opposite side of the Sun (see Figure \ref{c2aia}),
in the high corona the CME ceases its lateral expansion and decouples from its driven shock.
In contrast, the EUV wavefront speed is measured in the early stage when the wavefront is still coupled with the CME.
This may also weaken the correlation between the speeds of the EUV wavefront and the shock lateral expansion.

\subsection{Continuous Coupling of the EUV Wavefront and the CME boundary}\label{disCoupling}
The interpretation combining the wave and non-wave components of an EUV wave has become more popular,
which is supported by the observations of the decoupling of the EUV wavefront and the CME driver
\citep[e.g.,][]{PatsourakosV2009ApJ,ChengZO2012ApJ,LiberatoreLV2023ApJ}.
As detailed in Section \ref{coupling}, for the 2011 September 22 and 2013 November 7 events,
around the time when the EUV wavefront fades out at all altitudes,
the CME boundary is observed simultaneously at a similar position angle where the EUV wavefront vanishes
(see Column (6) of Table \ref{tabrange}).
This indicates that the EUV wavefront is coupled with the CME boundary loops for the whole propagation.
These two events are different from the other four events
in which the EUV wavefront decouples either before or when encountering a local loop system or coronal hole.
For the two events on 2011 September 22 and 2013 November 7,
in the path of the EUV wavefront propagation
there is also a local loop system in the low corona.
However, both the EUV wavefront and expanding CME structure traverse the loop system
and are coupled until the wavefront disappears.
During the lateral expansion of the CME,
the boundary loops in the low corona become more inclined and eventually lie on the solar surface,
while the loops are stretched straight in the high corona (e.g., above 2 \rsun{} as displayed in Figure \ref{c2aia}).
As a result, the CME extends laterally to as large as the travel distance of the EUV wavefront.
The two events suggest that an EUV wavefront can be continuously driven by the expanding CME and not be free during the whole propagation.
For the first time, we observe the continuous coupling of an EUV wavefront with the associated CME throughout the entire wavefront travel.
In this condition, the speed of the EUV wavefront is closely related to the lateral expansion of the CME
but not to the local fast magnetosonic speed,
which should be used with caution in coronal seismology.

\subsection{Connection with the EUV Wavefront and the Geometry of the CME-driven Shock}\label{disConnection}
Except for the 2017 September 10 and 2020 November 29 events,
after the EUV wavefront has faded out, the CME-driven shock continues propagating to the opposite side of the Sun,
which results in a large angular distance ($>70\degr$) between the shock flank and the EUV wavefront
(actually for the 2020 November 29 event the angular distance for the southward shock flank is also $>70\degr$).
With the angular distance $>70\degr$, geometrically the shock flank at $\sim$2 \rsun{}
cannot be linked to the EUV wavefront at $\sim$1 \rsun{} by a straight line.
Considering that the shock flank is convex and bulges to the propagation direction
(indicated by the arrow in Figure \ref{c2aia}),
when the shock flank is on the opposite side of the Sun, it can no longer connect with the EUV wavefront.
As suggested by models in which the fast magnetosonic speed increases with the altitude in the low corona
\citep[e.g.,][]{UchidaY1968SoPh,AfanasyevU2011SoPh,KwonV2017ApJ,MannV2023AA},
the inclination angle of a free wavefront in the low corona should decrease with the travel distance.
At a proper distance the inclination angle is reduced to nearly zero,
so the low-corona wavefront crashes onto the solar surface and eventually dissipates.
At the beginning a CME-driven shock is usually a closed dome that is connected to the EUV wavefront as the footprint
\citep[e.g.,][]{PatsourakosV2009ApJ,VeronigMK2010ApJ,LiZY2012ApJ,CunhaSilvaSF2018AA,MannV2023AA},
and in the late stage the shock has a sphere-like envelope
\citep[e.g.,][]{KwonV2017ApJ,LiuHZ2017ApJ,ZhuLK2021ApJ}.
In this study,
after the EUV wavefront has vanished and
the shock flank has traveled to the side opposite the eruption site,
the shock in the low corona dissipates and the closure of the sphere-like shock bubble begins to collapse.
This illustrates the final geometry of a large-scale CME-driven shock
when it has traveled to the opposite side and has disconnected from the corresponding EUV wavefront.
The final angular width of the CME-driven shock is obviously larger than that of the EUV wavefront.
This is different from the unique 2017 September 10 event,
where the shock is transmitted by polar coronal holes and
is connected with the EUV wavefront throughout the propagation \citep{HuLZ2019ApJ}.
Additionally, as shown in Figure \ref{c2aia}, all the shock flanks are far ahead of the CME boundaries,
which means that the laterally expanding shocks are freely propagating on the opposite side and are no longer being driven by their CMEs.
A CME-driven shock plays a key role in the longitudinal distribution of energetic particles
\citep[e.g.,][]{RouillardST2012ApJ,ParkIB2013ApJ,LarioKV2016ApJ,KouloumvakosPN2016ApJ,ZhuLK2018ApJ}.
Unveiling the geometry of a global CME-driven shock that has traveled to the other side of the Sun
contributes to the understanding of the release and distribution of energetic particles in the heliosphere.

\section{Conclusions}\label{conclusions}
We have selected and investigated six global EUV wave events associated with eruptions on the solar limb (listed in Table \ref{tablist}).
The eruptions are observed from the view of the Earth as halo CMEs
because the CME-driven shocks have laterally propagated to the other side opposite the eruption site.
The local morphology and kinematics of the EUV wavefront in the low corona from 1.05 to 1.25 \rsun{} have been analyzed.
By comparing the angular distance between the EUV wavefront and the CME boundary around the time when the wavefront fades out,
the coupling of the EUV wavefront and the CME driver has been discussed.
After the shock has propagated to the opposite side with a large angular distance
ahead of the position angle where the EUV wavefront vanished,
the CME-driven shock disconnects from the EUV wavefront,
which is observed in four events.
The key findings of this study are remarked as follows:

\begin{enumerate}
  \item The forward inclination of the primary EUV wavefront is observed
  in the early stage of the propagation for five of the six events,
  and the inclination angle ranges from 61\degr{} to 75\degr{}.
  The inclination is attributed to the local geometry of the forward inclined loops
  of the boundary of the expanding CME in the low corona,
  because the CME is coupled with the EUV wavefront in the early stage.

  \item The inclination angle of the EUV wavefront transmitted by coronal holes in two events
  is estimated to be $\sim$38\degr{} for the first time.
  The inclination can be explained by the increasing fast magnetosonic speed with the altitude in the low corona,
  given that the wavefront transmitted by a coronal hole is freely propagating.

  \item The large inclination angle of the transmitted wavefront can significantly
  affect the estimate of the actual speed which is perpendicular to the wavefront.
  This effect is up to $\sim40$\%{} and should be considered in the measurement of the actual speed.

  \item The wavefront speed is elevated by a local loop system
  only at lower altitudes below 1.15 \rsun{} in three of the six events,
  which results in distortion of the wavefront in the low corona.
  In addition to the wavefront inclination,
  this could make it difficult to estimate the actual speed
  of an on-disk EUV wavefront from single viewpoint observations.

  \item The average fast magnetosonic Mach numbers are in the range of 1.2--2.1
  at altitudes from 1.05 to 1.25 \rsun{} for four of the six events,
  and their corresponding average speeds are generally in the range of 300--500 \kmps{}.
  The Mach number larger than unity indicates that the EUV wavefront is a shock
  even at the coronal base in the initial stage when the EUV wavefront is still coupled with the CME.

  \item The average speed of the EUV wavefront has no obvious correlation with
  the average speed of the lateral expansion of the CME-driven shock.
  This may be because the EUV wavefront speed is measured in the early stage in the low corona
  while the shock speed is estimated in the later stage in the high corona.
  Additionally, in the early stage the EUV wavefront is still coupled with the expanding CME
  and in the late stage the shock has already decoupled from the CME driver.

  \item The EUV wavefront can be coupled with and driven by the CME during the entire propagation,
  which is observed in two of the six events.
  This suggests that the speed of the EUV wavefront is closely related to the lateral expansion of the CME,
  and thus the speed should be used cautiously in coronal seismology.

  \item In four of the six events, after the CME-driven shock has propagated to the opposite side of the Sun,
  it can be $>70$\degr{} away from the position angle where the EUV wavefront vanishes,
  which suggests that the shock is no longer connected with the EUV wavefront, losing its footprint in the low corona.
  After a long-distance propagation, the CME-driven shock in the low corona eventually crashes and dissipates on the solar surface.

\end{enumerate}

\begin{acknowledgments}
The research is supported by
the Strategic Priority Research Program of the Chinese Academy of Sciences (No. XDB0560000),
the National Natural Science Foundation of China
(Nos. 42004145, 42274201, 42150105, 12073032, and 42204176),
the National Key R\&D Program of China (Nos. 2022YFF0503800 and 2021YFA0718600),
and the Specialized Research Fund for State Key Laboratories of China.
C.C. is supported by the Research Foundation of Education Bureau of Hunan Province of China (No. 23B0593).
SDO is the first mission launched for NASA's Living With a Star Program.
SOHO is a project of international cooperation between ESA and NASA.
We acknowledge the use of SOHO/LASCO halo CME catalog,
which is generated and maintained at the CDAW Data Center by NASA
and the Catholic University of America in cooperation with the Naval Research Laboratory.
The MAS MHD simulation data are provided by Predictive Science Inc.,
and are available at \url{https://www.predsci.com/mhdweb/data_access.php}.

\end{acknowledgments}

\facilities{SDO, SOHO}
\software{SolarSoftWare (\citealt{FreelandH2012ssw}), JHelioviewer (\citealt{MuellerNF2017AA})}

\appendix
\section{{Local Loops in the Low Corona}} \label{pfss}

\begin{figure*}[ht!]
  \centering
  \includegraphics[width=0.62\textwidth]{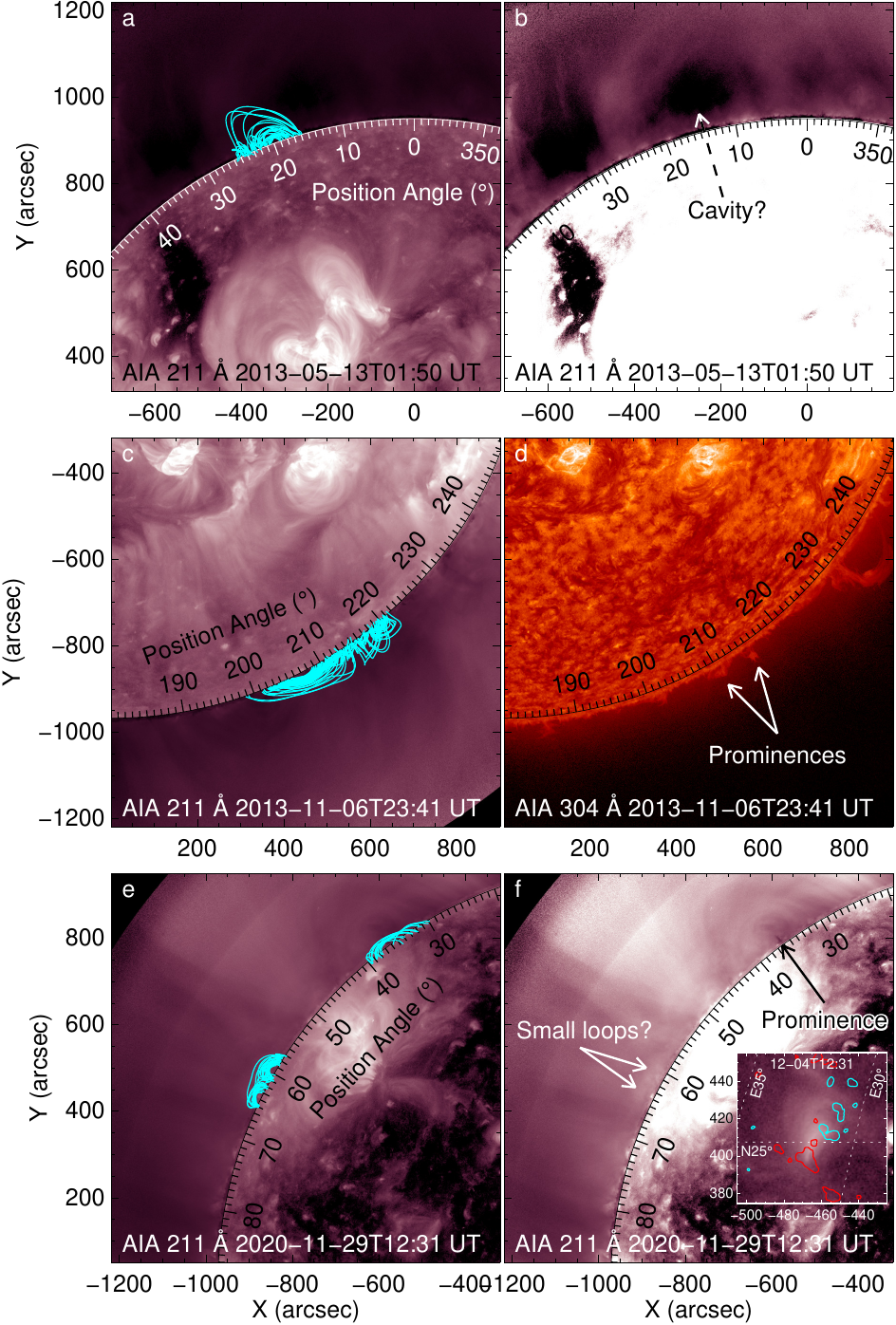} 
  \caption{\label{loops}{Cropped SDO/AIA images for the 2013 May 13 ((a)--(b)),
  2013 November 7 ((c)--(d)), and 2020 November 29 ((e)--(f)) events.
  In the left panels are AIA 211 {\AA} images,
  where the corona is enhanced using the SolarSoftWare routine \emph{aia\_rfilter}
  (see \url{http://aia.cfa.harvard.edu/rfilter.shtml} for details),
  and the cyan curves represent PFSS-extrapolated closed magnetic field lines by the \emph{pfss} package in SolarSoftWare
  (see \url{https://www.lmsal.com/~derosa/pfsspack/}).
  In (b) and (f) are the same images as in their corresponding left panels,
  but the contrast is adjusted to highlight the coronal features.
  In (d) is an AIA 304 \AA{} image corresponding to the 211 \AA{} image in (c).
  The arrow in (b) points to a plausible coronal cavity indicating the PFSS-extrapolated loops in (a).
  The arrows in (d) indicate two prominences (filaments) possibly in the PFSS-derived loop system in (c).
  The white arrows in (f) mark two small diffuse bright features
  corresponding to the PFSS-extrapolated loops near position angle 65\degr{} in (e);
  the black arrow in (f) denotes a small prominence possibly related to the loops near position angle 36\degr{} in (e).
  The inset overlaid in (f) is an AIA 211 \AA{} image on December 4
  taken after the limb near position angle 65\degr{} rotates to on-disk,
  which is superimposed with contours of positive (cyan) and negative (red) 20 Gauss line-of-sight magnetic fields
  from SDO/HMI \citep{SchouSB2012SoPh}.
  The ticks on the solar limb scale the position angle in degrees.}}
\end{figure*}

{Figure \ref{loops} shows the SDO/AIA images for the 2013 May 13, 2013 November 7, and 2020 November 29 events,
which are taken just before the onset of the corresponding flare.
These images illustrate the local loop systems in the low corona
that elevate the EUV wavefront speed as discussed in Sections \ref{kinematics} and \ref{disSpeed}.
The displayed magnetic field lines in Figure \ref{loops} are extrapolated with a potential-field source-surface (PFSS) model.
}

{For the 2013 May 13 event, as shown in Figure \ref{k20130513},
the wavefront speed is elevated between position angles $\sim 10\degr$ and $\sim 20\degr$.
In the range of the position angle, a possible coronal cavity can be seen in the AIA 211 \AA{} image in Figure \ref{loops}(b),
which may indicate a loop system
\citep[like][]{LiuON2012ApJ,Gibson2018LRSP}.
As shown in Figure \ref{loops}(a),
the PFSS model indeed reveals a loop system in the low corona that is consistent with the cavity.
However, the position-angle range of the PFSS-extrapolated loops does not correspond to the range of the 211 \AA{} cavity,
where no closed magnetic field lines are obtained below a position angle $\sim$15\degr{}.
This may be due to (1) the magnetogram data on the solar limb used by the PFSS model are not real-time and are taken more than 10 days ago,
and (2) the observational uncertainty of magnetogram data in the polar regions impedes the determination of the actual magnetic field structures.
}

{For the 2013 November 7 event, as displayed in Figure \ref{k20131107},
the wavefront speed is locally elevated near position angle 215\degr{}.
A group of closed magnetic field lines are extrapolated near the position by the PFSS model,
and are spatially coincident with two small prominences, as illustrated in Figure \ref{loops}(c)--(d).
A prominence (filament) is a cool, dense structure with twisted magnetic field lines,
which is usually embedded in overlying magnetic loops \citep[see the review by][]{Gibson2018LRSP}.
}

{For the 2020 November 29 event, as shown in Figure \ref{k20201129},
the wavefront speed at low altitudes increases near position angle 65\degr{}.
A set of closed magnetic field loops are produced by the PFSS model near this position,
which is illustrated in Figure \ref{loops}(e).
In the contrast-enhanced AIA 211 \AA{} image in Figure \ref{loops}(f),
there are two small diffuse bright structures corresponding to the PFSS-extrapolated loops,
which suggests these loops are from a bright point with bipolar magnetic fields
\citep[refer to a review by][]{Madjarska2019LRSP}.
After the east solar limb rotates to on-disk five days later,
we indeed see a bright point with bipolar magnetic fields in the corresponding region
(see the inset AIA 211 \AA{} image in Figure \ref{loops}(f)).
There is a small prominence indicated by the black arrow in Figure \ref{loops}(f),
which is probably below another set of loops corresponding to the PFSS-extrapolated closed field lines
between position angles 30\degr{} and 40\degr{} in Figure \ref{loops}(e).
The feature in the enhanced corona above the prominence in the 211 \AA{} image also exhibits a loop-like structure,
although the height is not as large as that of the PFSS-extrapolated loops.
These loops may explain the increased MHD-derived fast magnetosonic speed at position angles below 40\degr{}
as shown in Figure \ref{k20201129} in Section \ref{kinematics}.
}

{The PFSS-extrapolated loops displayed in Figure \ref{loops}
are selected by limiting the start points of the field lines to be within a small range of longitudes on the limb
and of latitudes where the EUV wavefront speed is elevated in the low corona for each of the three events.
The maximum height of the field lines is not limited in the selection,
but the extrapolated loops are all in the low corona.
Although the magnetogram data on the limb used by the PFSS model cannot be acquired in real-time,
the extrapolated low-corona loops are consistent with the low-corona elevation of the wavefront speed
as discussed in Section \ref{kinematics} and \ref{disSpeed}.
}


\end{CJK*}
\end{document}